# New scaling laws for pinning force density in superconductors

Evgeny F. Talantsev[1,2]

[1]M.N. Miheev Institute of Metal Physics, Ural Branch, Russian Academy of Sciences, 18, S. Kovalevskoy St., Ekaterinburg, 620108, Russia

[2]NANOTECH Centre, Ural Federal University, 19 Mira St., Ekaterinburg, 620002, Russia

## Abstract

Since the report by Fietz and Webb (1968 *Phys. Rev.* **178** 657), who considered the pinning force density, $\overrightarrow{F_p} = \overrightarrow{J_c} \times \vec{B}$ (where, $J_c$ is the critical current density and $B$ is applied magnetic flux density), in isotropic superconductors as a unique function of reduced applied magnetic flux density, $\frac{B}{B_{c2}}$ (where $B_{c2}$ is the upper critical field), $\left|\overrightarrow{F_p}\right|$ has been scaled based on $\frac{B}{B_{c2}}$ ratio, for which there is widely used scaling law of $\left|\overrightarrow{F_p}(B)\right| = F_{p,max} \cdot \left(\frac{B}{B_{c2}}\right)^p \cdot \left(1 - \frac{B}{B_{c2}}\right)^q$, where $F_{p,max}$, $B_{c2}$, $p$, and $q$ are free-fitting parameters, proposed by Kramer (1973 *J. Appl. Phys.* **44** 1360) and Dew-Hughes (1974 *Phil. Mag.* **30** 293). To describe $\left|\overrightarrow{F_p}(B)\right|$ in high-temperature superconductors, Kramer-Dew-Hughes scaling law was modified by (a) an assumption of the angular dependence of all free-fitting parameters on the rotation angle θ and (b) by the replacement of the upper critical field, $B_{c2}$, by the irreversibility field, $B_{irr}(\theta)$. Here we note that the pinning force density is also a function of critical current density and, thus, $\left|\overrightarrow{F_p}(J_c)\right|$ scaling law should exist. In an attempt to reveal this law, we considered full $\left|\overrightarrow{F_p}(B, J_c)\right|$ function and reported that there are three distinctive characteristic ranges of $\left(\frac{B}{B_{c2}}, \frac{J_c}{J_c(sf)}\right)$ (where $J_c(sf)$ is the self-field critical current density) on which $\left|\overrightarrow{F_p}(B, J_c)\right|$ can be splatted. Several new scaling laws for $\left|\overrightarrow{F_p}(J_c)\right|$ were proposed, discussed and applied to approximate $\left|\overrightarrow{F_p}\right|$ in $MgB_2$, NdFeAs(O,F), REBCO, and near-room temperature superconducting superhydrides $(La,Y)H_{10}$ and $YH_6$. We pointed out that the general scaling law for the pinning force density is on the quest.

## New scaling laws for pinning force density in superconductors

### 1. Introduction

In 1962 Philip W. Anderson wrote [1]: "A major difficulty in understanding hard superconductors has been the appreciable temperature dependence of critical currents and fields at temperatures as low as $0.1 \cdot T_c$. None of the properties of the bulk superconducting state vary noticeably at this temperature. …

…We shall show that this … can be explained by assuming that the mechanism of flux creep is thermally activated motion of bundles of flux lines, aided by the Lorentz force $\vec{J} \otimes \vec{B}$, over free energy barriers coming from the pinning effect of inhomogeneities, strains, dislocations, or other physical defects."

Since then, the primary idea that the origin of the electric power dissipation in type-II superconductors is the Abrikosov's vortices dissipative movement under the Lorentz force [2] has been unconditionally accepted.

Based on this concept, superconducting materials can be characterized by the pinning force density, $\overrightarrow{F_p}$, defined as:

$$\overrightarrow{F_p} = \overrightarrow{J_c} \otimes \vec{B} \tag{1}$$

where $J_c$ is the critical current density and $B$ is the applied magnetic flux density. In this paper, we performed a thorough analysis of the basic properties of the amplitude of this value $\left|\overrightarrow{F_p}\right| = \left|\overrightarrow{J_c} \otimes \vec{B}\right|$ in thin film superconductors, including $MgB_2$, cuprates, and pnictides.

Fietz and Webb [3] were the first who proposed a scaling law for this value:

$$\left|\overrightarrow{F_p}\right| = \left|\overrightarrow{F_p}(B)\right| = K \times k(\kappa) \times B_{c2}^{5/2} \times g\left(\frac{B}{B_{c2}}\right) \tag{2}$$

where $K$ is a numerical multiplicative pre-factor, $k(x)$ is a function where independent variable is the Ginzburg-Landau parameter $\kappa$, $B_{c2}$ is the upper critical field and $g(x)$ is a function of reduced applied magnetic field. Several years later, Kramer [4] and Dew-Hughes [5] developed the most widely used expression for the $|\overrightarrow{F_p}|$:

$$|\overrightarrow{F_p}(B)| = F_{p,max} \times \left(\frac{B}{B_{c2}}\right)^p \times \left(1 - \frac{B}{B_{c2}}\right)^q \tag{3}$$

where $F_{p,max}$, $B_{c2}$, $p$, and $q$ are free-fitting parameters. It should mention that Eq. 3 is in use in its normalized form of [6]:

$$|\overrightarrow{f_{p,n}}(B)| = \frac{(p+q)^{p+q}}{p^p \cdot q^q} \times \left(\frac{B}{B_{c2}}\right)^p \times \left(1 - \frac{B}{B_{c2}}\right)^q \tag{4}$$

For $Nb_3Sn$-based alloys, Eq. 3 was upgraded [7-11] to account the temperature and the strain dependence of the $|\overrightarrow{F_p}|$ by the introduction of several multiplicative terms in Eq. 3:

$$|\overrightarrow{F_p}(B,T,\varepsilon)| = C \times (1 - a \times |\varepsilon - \varepsilon_m|^{1.7})^s \times \left(\frac{B}{B_{c2}(\varepsilon,T)}\right)^p \times \left(1 - \frac{B}{B_{c2}(\varepsilon,T)}\right)^q \times \left(1 - \left(\frac{T}{T_c}\right)^2\right)^\mu \times \left(1 - \left(\frac{T}{T_c}\right)^{1.52}\right)^{\eta-\mu} \tag{5}$$

where $C$, $a$, μ, η, $\varepsilon_m$, $s$, $T_c$ are free-fitting parameters and $B_{c2}(\varepsilon,T)$ is given by different function. On one hand Eq. 5 is applicable only for $Nb_3Sn$ and only at moderate strain level [7-11]. On other hand the upgraded equation 5 is a fitting function constructed in terms of the general flavor of the two-fluid model, where each new parameter is added to un-upgrade equation through a multiplicative term of:

$$V(P) = V_0 \times \left(1 - \left(\frac{P}{P_{max}}\right)^m\right)^n \tag{6}$$

where $V(P)$ is the fitted value with new variable $P$, $V_0$, $P_{max}$, $m$, $n$ are free fitting parameters, $P_{max}$ is maximal value for the new parameter (it is chosen within given model, for instance, temperature dependences of the superconducting parameters are generally utilized $P_{max} = T_c$, which is the maximum temperature at which the superconducting state does exist).

In this regard, there is another often chosen form for the new fitting term is based on exponential function [12], from which we can mention the scaling law proposed by Fietz *et al*. [13] for Nb-Zr superconductors:

$$J_c(B,T) = a_0 \times exp^{-\left(\frac{B}{B_0}\right)} + C_0 \quad (7)$$

where $a_0$, $B_0$ and $C_0$ are free-fitting parameters of the model. The temperature-dependent fitting term also can be represented by an exponential function, as it was proposed for the second-generation high-temperature superconducting wires (2G-wires) by Senatore *et al* [14]:

$$J_c(B,T) = J_c(B=0, T=0) \times exp^{-\left(\frac{T}{T^*}\right)} \times B^{-\alpha} \quad (8)$$

where $J_c(B=0, T=0)$, $T^*$ and $\alpha$ are free-fitting parameters ($0.2 \leq \alpha \leq 0.9$ [14]).

It should be noted that Eq. 8 is internally incorrect because the units of the right hand of the equation are not $A/m^2$ (which is current density units). This mistake is becoming widely spread, because recently Francis *et al* [15] utilized Eq. 8 (which is Eq. 1 in Ref. 15) for the analysis of nanostructured coated conductors.

Overall, for $MgB_2$, cuprates, and iron-based superconductors (IBS), different expressions for the $\left|\overrightarrow{F_p}(B)\right|$ were proposed [6,12-19] (extended review for REBCO has given by Jirsa *et al* [6]). For these compounds, due to their anisotropic nature, all free-fitting parameters of Eqs. 3,4 are angular dependent. Also, due to the relatively wide transition width in these compounds, the primary ratio $\left(\frac{B}{B_{c2}}\right)$ in scaling laws (Eqs. 3,4) is often replaced by $\left(\frac{B}{B_{irr}(\theta)}\right)$, where $B_{irr}(\theta)$ is the irreversibility field.

Despite some differences, a primary assumption of all known scaling laws [3-19] is that $\left|\overrightarrow{F_p}\right|$ should have the applied magnetic field, $B$, as a primary variable (Eqs. 1-5). Thus, all known scaling law have developed for $\left|\overrightarrow{F_p}(B)\right|$ values.

In this paper, we pointed out that $\left|\overrightarrow{F_p}\right|$ (Eq. 1) is also a function of $J_c$. This fact was, for somehow, never considered. And thus, $\left|\overrightarrow{F_p}(J_c)\right|$ scaling law should also exist. Based on this primary idea, we consider full $\left|\overrightarrow{F_p}(J_c, B)\right|$ function and propose potential scaling laws for $\left|\overrightarrow{F_p}(J_c)\right|$ in thin film superconductors which reflect different pinning characteristics of the materials. We limited our paper by the case when an external magnetic field, $B$, is applied, in the perpendicular direction, to the film's large surface, i.e., for the field angle θ = 0° [20].

## 2. Problem description

Fundamental problem for existing approach to scale $\left|\overrightarrow{F_p}\right|$ as a solely function of applied magnetic field, *B*, which we found in this paper is shown in Figure 1, where $\left|\overrightarrow{F_p}(J_c, B, T = 4.2\ K)\right|$ curve for $MgB_2$ thin film is shown. It should be noted that raw $J_c(B, T, \theta = 0°)$ data for this films was reported by Zheng *et al* [20], and $J_c(B, T = 4.2\ K, \theta = 0°)$ dataset is shown in Figure 2(a). Also, it needs to mention that in all 3D representations of $\left|\overrightarrow{F_p}(J_c, B)\right|$, we used the same axes arrangement, where X-axis is $J_c$, Y-axis is *B*, and Z-axis is $\left|\overrightarrow{F_p}(J_c, B)\right|$.

It can be seen in Figure 1, that $\left|\overrightarrow{F_p}(B)\right|$ curve (green), which is traditionally used for the pinning force analysis, represents a projection of 3D $\left|\overrightarrow{F_p}(J_c, B)\right|$ curve (red) onto the X = 0 plane. It should be stressed that at low magnetic field, $B \lesssim \frac{1}{4} B_{c2}$, entire $\left|\overrightarrow{F_p}(J_c, B)\right|$ curve (red) is directed nearly in perpendicular direction to the X = 0 plane. For this reason, the projection (green curve) of this part of $\left|\overrightarrow{F_p}(J_c, B)\right|$ curve onto the X = 0 plane cannot be considered as a correct representation of the entire pinning force density $\left|\overrightarrow{F_p}(J_c, B)\right|$ curve.

This is primary reason why widely accepted deduction [6-19,22,23] of *p*-parameter of the Eq. 3, which is primary value in Kramer-Dew-Hughes scaling laws [2,3] at $\frac{B}{B_{c2}} \lesssim \frac{1}{4}$:

$$\left|\overrightarrow{F_p}(B)\right|_{B \lesssim \frac{1}{4} B_{c2}} = F_{p,max} \times \left(\frac{B}{B_{c2}}\right)^p \times \left(1 - \frac{B}{B_{c2}}\right)^q \Big|_{B \lesssim \frac{1}{4} B_{c2}} \rightarrow\ F_{p,max} \times \left(\frac{B}{B_{c2}}\right)^p, \qquad (9)$$

cannot be considered to exhibit a sustainable meaning.

Geometrical representation showed in Figure 1 (where the pinning force density line is directed in nearly orthogonal direction to the X = 0 plane) also explains, why Eq. 9 cannot be used to extract $J_c(B)$ at low applied magnetic field, as well the self-field critical current density, $J_c(B = sf)$ (when applied magnetic field $B$ = 0 T), from the pinning force density:

$$\left|\overrightarrow{F_p}(B)\right|_{B \lesssim \frac{1}{4}B_{c2}} \cong F_{p,max} \times \left(\frac{B}{B_{c2}}\right)^p = J_c \times B, \quad (10)$$

$$J_c\left(B \lesssim \frac{1}{4}B_{c2}\right) = \frac{F_{p,max}}{(B_{c2})^p} \times \frac{1}{B^{1-p}}, \quad (11)$$

where $p < 1$, and, thus:

$$\lim_{B \to 0} J_c(B) = \infty \quad (12)$$

which contradicts with experimental observations.

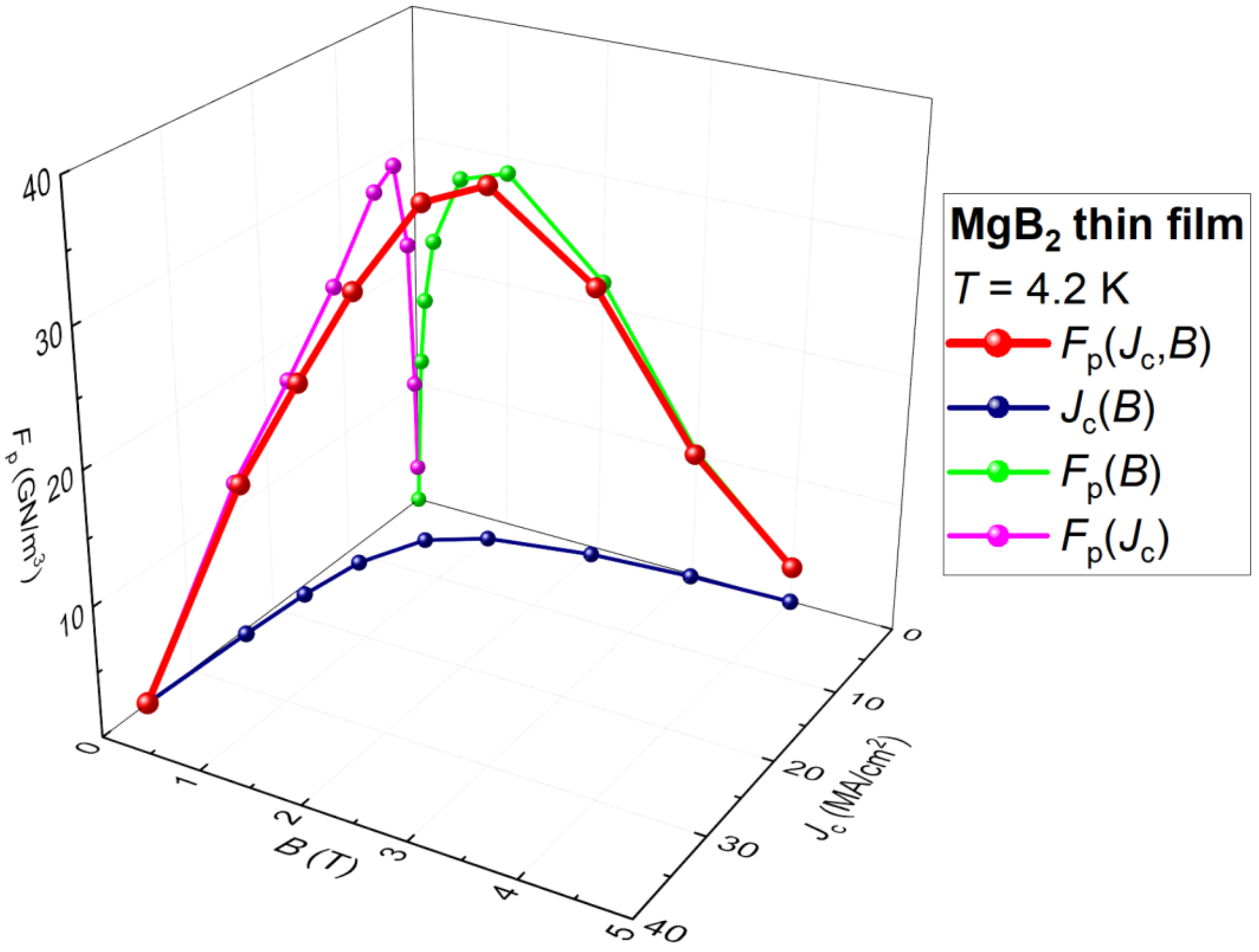

**Figure 1.** Three-dimensional (3D) representation of the pinning force density, $\left|\overrightarrow{F_p}(J_c, B, T = 4.2\ K)\right|$, for $MgB_2$ thin film. Raw $J_c(B, T = 4.2\ K)$ data was reported by Zheng *et al* [21].

It should be also mentioned that existing theory of Abrikosov's flux pinning [5] cab be used to derive solid theoretical values for $p$-parameter in Eq. 3 for certain types of pinning mechanisms, defects dimensionality, vortexes pinning length, etc. (see, for instance, Ref. 5). However, it can be seen from all above, this parameter cannot be accepted to be sustainable value which characterizes the pinning force density (Eq. 1) in superconductors.

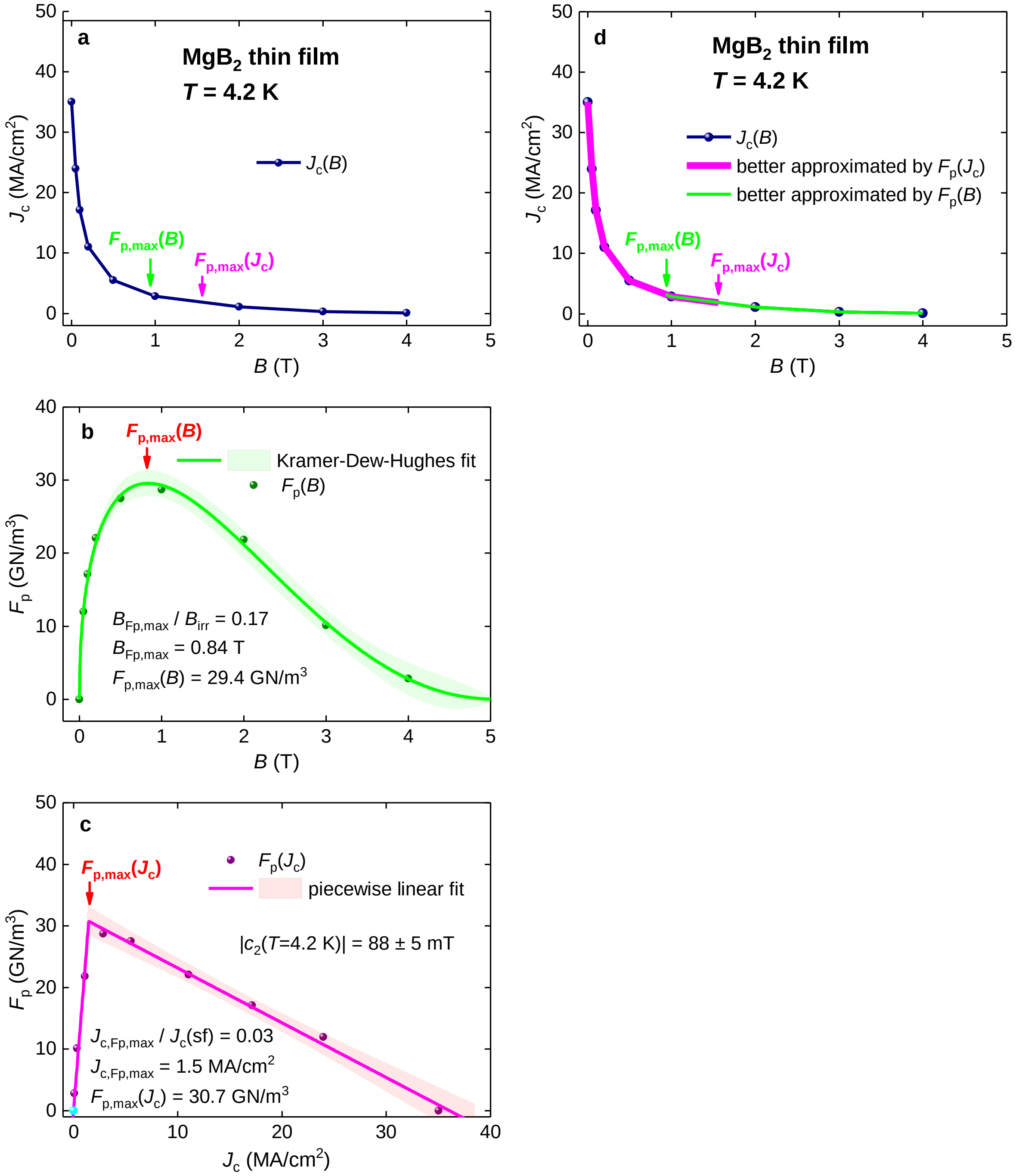

**Figure 2.** Projections of the $\left|\overrightarrow{F_p}(J_c, B, T = 4.2\ K, \theta = 0°)\right|$ curve for $MgB_2$ thin film into three orthogonal planes. Data fit for **b** and **c** panels are shown. **a** – the projection into the $F_p$=0 plane; **b** – the projection into the $J_c$=0 plane and data fit to Kramer-Dew-Hughes (equation 3); and **c** – the projection into the $B$=0 plane and data fit to linear piecewise model (equation 13), cyan ball at the origin indicates $\left|\overrightarrow{F_p}(J_c)\right|$ at $B = B_{c2}$; **d** – the projection into the $F_p$=0 plane, where two overlapped branches are shown (see details in the text). Raw $J_c(B, T = 4.2\ K, \theta = 0°)$ data was reported by Zheng *et al* [21]. 95% confidence bands are shown by shadow areas.

To demonstrate that $\left|\overrightarrow{F_p}(B,J_c)\right|$ dataset for the $MgB_2$ thin film (showed in Figure 1) represents typical dataset for superconducting films, we fitted the projection of this dataset into the X = 0 plane, i.e. $\left|\overrightarrow{F_p}(B)\right|$, to Eq. 3 in Figure 2(b). Deduced parameters $p = 0.45 \pm 0.03 \cong 0.5$ and $q = 2.3 \pm 0.5 \cong 2$ certainly represent typical characteristic values which are often deduced from the $\left|\overrightarrow{F_p}(B)\right|$ fits to Eq. 3 [11,23,24].

It is crucial to note, that this set of parameters, $p = 0.5$ and $q = 2$, was listed by Dew-Hughes in his Table (Ref. 5) and this set of parameters implies that the flux pinning is due to the grain boundaries pinning [5]. However, films fabricated by Zheng *et al* [21] are epitaxial films on single crystal substrates and, thus, there is no expectation that grain boundaries are exhibited in these $MgB_2$ thin films.

Considering another projection of 3D $\left|\overrightarrow{F_p}(J_c,B)\right|$ curve into the Z = 0 plane, which is $J_c(B)$ (royal curve in Fig. 1 and Fig. 2,a), we should mention that $J_c(B)$ is largely distant from the $\left|\overrightarrow{F_p}(J_c,B)\right|$ curve, except two points of $J_c(B=0)$ and of $J_c(B=B_{c2})$, where the $\left|\overrightarrow{F_p}(J_c,B)\right|$ curve intersects the plane (Fig. 1). This implies that $J_c(B)$ projection represents the most distorted projection of 3D $\left|\overrightarrow{F_p}(J_c,B)\right|$ curve from three orthogonal projections of this curve. Thus, if one assumes that primary idea expressed by Anderson [1,2] is correct (i.e. that equation 1 describes upper limit for the dissipative-free transport current flow), then the use of $J_c(B)$ curve for the analysis of pinning properties of materials represents the most inappropriate choice of raw experimental data for the analysis.

If considered above two projections of 3D $\left|\overrightarrow{F_p}(J_c,B)\right|$ curve (i.e., $\left|\overrightarrow{F_p}(B)\right|$ and $J_c(B)$) are in a wide use to characterise superconducting materials, the third projection, $\left|\overrightarrow{F_p}(J_c)\right|$ (magenta curve in Figs.1,2(c)), from the best author's knowledge is introducing herein. $\left|\overrightarrow{F_p}(J_c)\right|$ exhibits nearly linear dependence on both sides from the $F_{p,max}(J_c)$ point (Figs.

1,2(c)). Based on this, $F_p(J_c)$ for this $MgB_2$ thin film can be approximated by a piecewise function of two linear functions:

$$F_p(J_c) = \theta(J_c \le J_{c,Fp,max}) \times (c_1 \times J_c + f_1) + \theta(J_c \ge J_{c,Fp,max}) \times (c_2 \times J_c + f_2) \quad (13)$$

where $c_1, f_1, c_2, f_2$ are free-fitting parameters and $\theta(x)$ is the Heaviside function. The fit of $F_p(J_c)$ to Eq. 13 is shown in Fig. 2(c). Based on the deduced ratio of (Fig. 2(c)):

$$\frac{J_{c,Fp,max}}{J_c(sf)} = 0.03 \quad (14)$$

one can conclude that nearly full $F_p(J_c)$ dependence, i.e. the range is:

$$0.03 \times J_c(sf) \le J_c \le J_c(sf), \quad (15)$$

can be approximated by a simple linear equation:

$$F_p(J_c) \equiv J_c(B) \times B = (c_2 \times J_c(B) + f_2) \quad (16)$$

$$J_c(B) = \frac{f_2}{B - c_2} = \frac{f_2}{B + |c_2|} = \frac{\frac{f_2}{|c_2|}}{1 + \frac{B}{|c_2|}} = \frac{J_c(B=0)}{1 + \frac{B}{|c_2|}} \quad (17)$$

Surprisingly enough, Eq. 17 is well known Kim model equation [24,25]. Our derivation is solely based on 3D graphical representation of the $|\overrightarrow{F_p}(J_c, B)|$ dataset (Fig. 1).

To answer a question, what is the physical meaning of the $|c_2|$ parameter in Eq. 17, we first should mention that this parameter has the unit of magnetic flux density (i.e. Tesla), and that this parameter is neither $B_{c2}$, nor $B_{irr}$, because its absolute value for the $MgB_2$ film is $|c_2(T = 4.2\ K)| = 88 \pm 5\ mT$ (Fig. 2) is by about two orders of magnitude below the upper critical field value. However, the latter value, which can be called the pinning field, $|c_2(T)|$, is comparable with $B_{c1}(T = 4.2\ K)$ (see, for instance, [27-30]), which indicates that $|c_2(T)|$ is related to low- and middle-field part of $|\overrightarrow{F_p}(J_c, B)|$ curve.

Considering that $B_{c1}(T)$ is the magnetic field at which Abrikosov's vortexes have thermodynamic preference to exhibit in the superconductor, it is useful to quantify the deduced pinning field, $|c_2(T)|$, vs the fundamental lower critical field of the material (at low-and middle-ranges of applied magnetic field) by a factor of $|c_2(T)|$ enhancement over $B_{c1}(T)$:

$$\varepsilon = \frac{|c_2(T)|}{B_{c1}(T)} \qquad (18)$$

Considering that the lower critical field is given by [31]:

$$B_{c1}(T) = \frac{\phi_0}{4\pi} \times \frac{ln\left(1+\sqrt{2}\kappa(T)\right)}{\lambda^2(T)}, \qquad (19)$$

and the self-field critical current density [32]:

$$J_c(B = 0) = \frac{\phi_0}{4\pi\mu_0} \times \frac{ln\left(1+\sqrt{2}\kappa(T)\right)}{\lambda^3(T)} \qquad (20)$$

one can obtain: calculate the lower limit of the pinning force density, $F_{p,lower\ limit}(T)$, which can be achievable in the pinning-free thin film:

$$B_{c1}(T) = \left[\frac{\phi_0}{4\pi} \times ln\left(1 + \sqrt{2} \times \kappa(T)\right)\right]^{\frac{1}{3}} \times \left(\mu_0 \times J_c(sf, T)\right)^{\frac{2}{3}} \qquad (21)$$

Utilizing reported $J_c(B = 0, T = 4.2\ K) = 35\ \ MA/cm^2$ [20] and $\kappa(T) = 26$ [33], one can calculate $B_{c1}(T = 4.2\ K) = 49\ mT$, which implies that the pinning field can be characterized by the enhancement factor $\varepsilon(T = 4.2\ K) = 1.8$.

We should be clear, that the applied magnetic field, for which the $|c_2(T = 4.2\ K)| = 0.088\ \pm 0.005\ T$ field was deduced, covers a wide range:

$$5 \times 10^{-4}\ T \leq B \leq 1.0\ T \qquad (22)$$

and the deduce values (i.e. $|c_2|$ and $\varepsilon$) characterize the pinning properties of given $MgB_2$ film within a wide range of applied magnetic field (i.e. low and middle applied field ranges).

From the conceptual point of view, $|c_2(T)|$ field can be interpreted as so-called matching field, which defines as the field at which each defect in the material holds one Abrikosov's vortex. Despite intensive attempts (which were mainly focused on heavy ion irradiation of cuprates, where ion damaging tracks can be prepared in the form of continuous lines and, thus, the matching field can be calculated) [34-36], the matching field has been never reliably extracted from experimental $J_c(B,T)$ data.

If this interpretation is correct, then it can be useful to introduce a characteristic hexagonal lattice parameter, $d_{hexagonal}(T)$, between Abrikosov's vortices which corresponds to the $|c_2|$ field in the assumption of hexagonal vortices lattice [37]:

$$d_{hexagonal}(T) = \left(\frac{\phi_0}{\frac{1}{2}\sqrt{3}|c_2(T)|}\right)^{\frac{1}{2}}, \tag{23}$$

For the given $MgB_2$ film, the equivalent lattice parameter is $d_{hexagonal}(T = 4.2\ K) = 165\ nm$, which indicates that this film is nearly pinning sites free.

At high applied magnetic field range the $\left|\overrightarrow{F_p}(J_c, B)\right|$ curve (red) is directed nearly in perpendicular direction to the Y = 0 plane. For this reason, the projection (magenta curve) of this part of $\left|\overrightarrow{F_p}(J_c, B)\right|$ curve into the Y = 0 plane cannot be considered as accurate approximation for the 3D pinning force density $\left|\overrightarrow{F_p}(J_c, B)\right|$ curve.

However, essential difference of the $\left|\overrightarrow{F_p}(J_c)\right|$ projection from its orthogonal counterparts and the 3D $\left|\overrightarrow{F_p}(J_c, B)\right|$ curve itself is that the $\left|\overrightarrow{F_p}(J_c)\right|$ dataset always includes the origin point in total dataset. Truly, by the definition $\left|\overrightarrow{F_p}(J_c, B)\right| = 0$ at $J_c = 0$ at any $B = B_{c2}$ (even for the case when $B_{c2}$ value is unknown). In Figure 2(c) this origin point is depicted by cyan colour.

In overall, Figure 1 shows that the 3D $\left|\overrightarrow{F_p}(J_c, B)\right|$ curve at low applied magnetic field is better approximated by $\left|\overrightarrow{F_p}(J_c)\right|$ projection, while at high applied magnetic field the curve is better approximated by $\left|\overrightarrow{F_p}(B)\right|$. These two curves are overlapped at some middle range of the applied magnetic field. The limiting values for overlapping range can be chosen to correspond to the field values at $\left|\vec{F}_{p,max}(B)\right|$ and $\left|\vec{F}_{p,max}(J_c)\right|$, as it shows in Figure 2(d).

Thus, we can propose to split a $\left|\overrightarrow{F_p}(J_c, B)\right|$ curve in three distinctive ranges:

1. low reduced applied magnetic field range, where $\left|\overrightarrow{F_p}(J_c, B)\right|$ can be approximated by $\left|\overrightarrow{F_p}(J_c)\right|$ projection;

2. high reduced applied magnetic field range, where $\left|\overrightarrow{F_p}(J_c, B)\right|$ can be approximated by $\left|\overrightarrow{F_p}(B)\right|$ projection;

3. middle reduced applied magnetic field, where $\left|\overrightarrow{F_p}(J_c)\right|$ and $\left|\overrightarrow{F_p}(B)\right|$ branches are overlapped and an approximation can be achieved by a constructing some new function, where both branches are presented with some weights.

In next Sections we showed the usefulness of this approach to analyse $\left|\overrightarrow{F_p}(J_c, B)\right|$ data for thin films of $MgB_2$, NdFeAs(O,F), REBCO, and near-room temperature superconducting superhydrides (La,Y)$H_{10}$ and $YH_6$. It should be mentioned that we considered only transport current datasets to avoid additional complication with the conversion of the magnetization data into critical current density.

## 3. Results and Discussion

### 3.1. $MgB_2$

The pinning force density $\left|\overrightarrow{F_p}(B, T)\right|$ in pure and doped $MgB_2$ films are studied in detail [38-40]. Here we have chosen for the analysis $\left|\overrightarrow{F_p}(B, J_c, T)\right|$ dataset reported by Zheng *et al* [21], who fabricated high-quality $MgB_2$ films by hybrid physical-chemical vapor deposition on two types of silicon carbide single crystal substrates (4H-SiC and 6H-SiC) and reported transport $J_c(B, T)$ datasets for the film deposited on a 6H-SiC substrate. The $\left|\overrightarrow{F_p}(J_c, B, T = 4.2\ K)\right|$ dataset was already analysed in Figures 1,2 and in Figure 3 other $\left|\overrightarrow{F_p}(J_c, B, T)\right|$ datasets are shown together with fits to Eq. 13.

There are three findings revealed by the $\left|\overrightarrow{F_p}(J_c)\right|$ data fits to Eq. 13:

1. A linear dependence of the $\left|\overrightarrow{F_p}(J_c)\right|$ vs $J_c$ remains at all temperatures (Fig. 3).

2. The $J_c$ range, where this linear dependence exhibits, covers nearly full $J_c$ range:

$$0.04 \times J_c(sf,T) \leq J_c(B,T) \leq J_c(sf,T). \quad (24)$$

3. The pinning field enhancement factor, $\varepsilon(T)$ (Eq. 18), remains nearly the same for full temperature range (Figure 4):

$$1.45 \leq \varepsilon(T) \leq 1.80 \text{ (for } 4.2\ K \leq T \leq 35\ K) \quad (25)$$

This implies that the characteristic field, $|c_2(T)|$ (Eq. 13), for the $MgB_2$ film can be approximately scaled as the lower critical field, $B_{c1}(T)$, for this material.

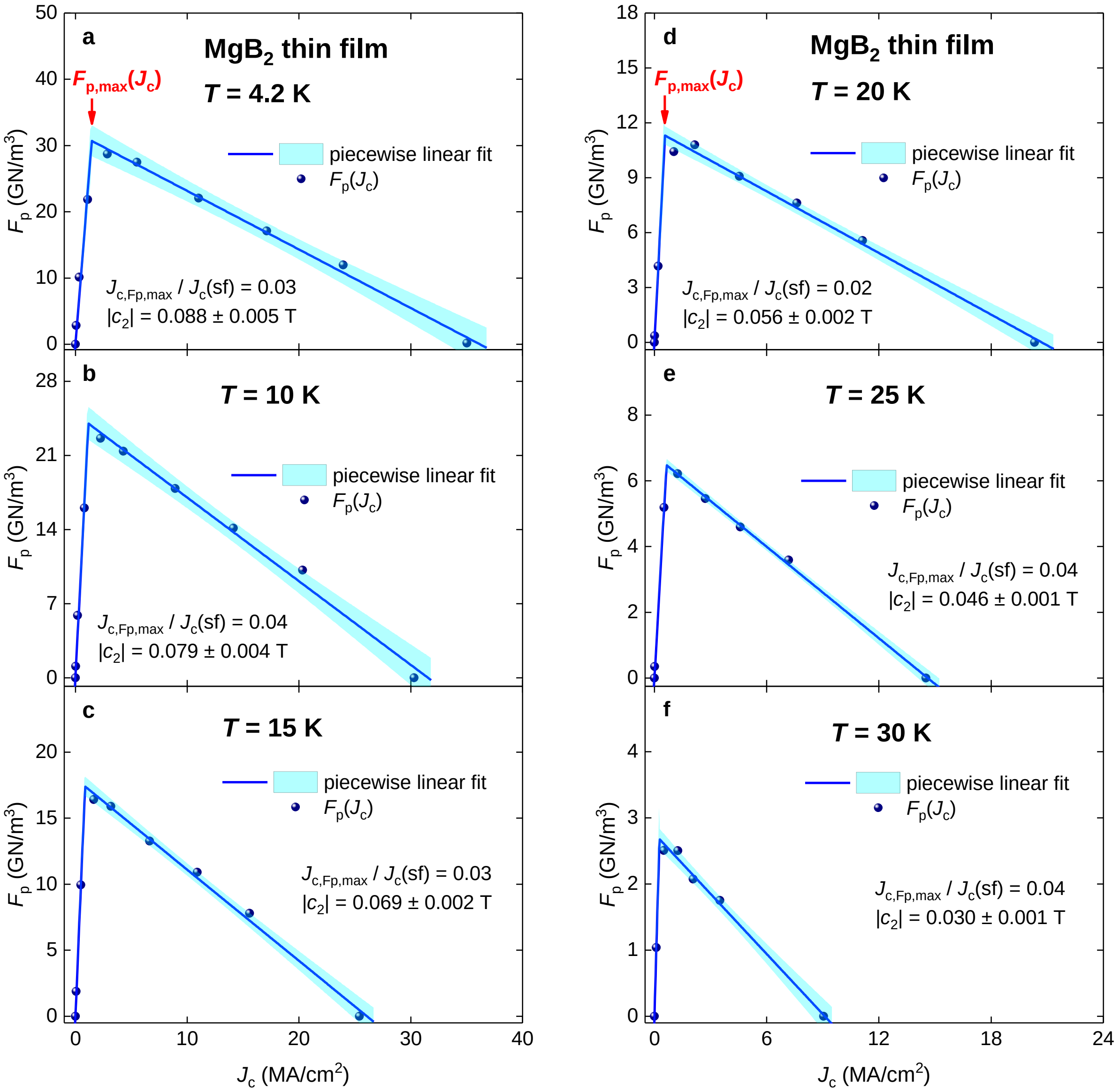

**Figure 3.** $\left|\overrightarrow{F_p}(J_c,T,\theta=0°)\right|$ curves and data fits to Eq. 13 for $MgB_2$ thin film deposited on a 6H-SiC single crystal substrate for temperature range 4.2-30 K (a-f). Raw transport $J_c(B,T,\theta=0°)$ data was reported by Zheng *et al* [21]. Goodness of fit for all panels is better than $R$ = 0.985. 95% confidence bands are shown by shadow areas.

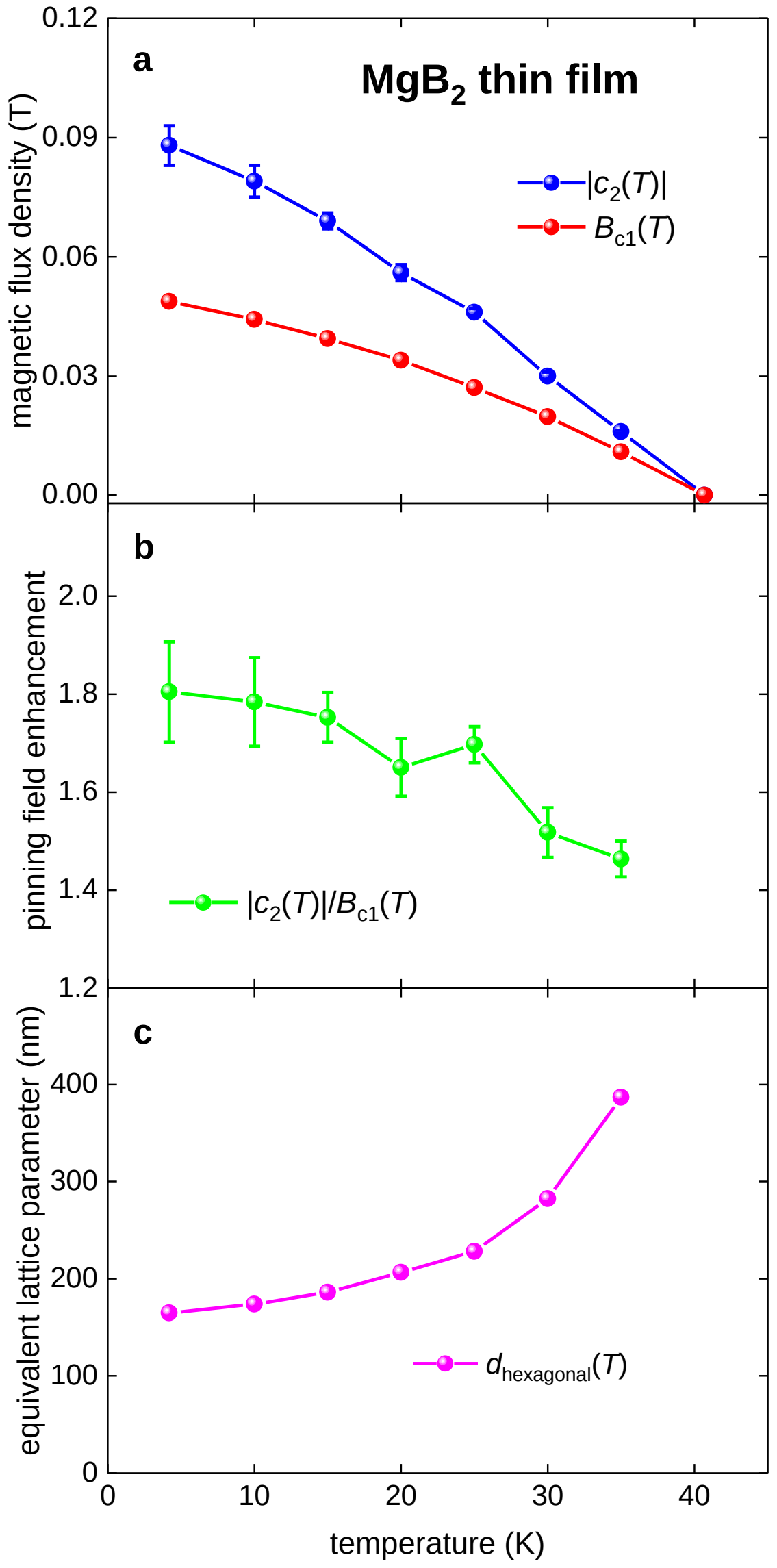

**Figure 4. a** - Deduced free-fitting parameter $|c_2(T)|$ (Eq. 13) and $B_{c1}(T)$ calculated by Eq. 21 for $MgB_2$ thin film deposited on a 6H-SiC single crystal substrate for which raw transport $J_c(B, T, \theta = 0°)$ data was reported by Zheng *et al* [21]. **b** – The pining field enhancement factor $\varepsilon(T)$ (Eq. 18) for the $MgB_2$ film. **c** – Equivalent hexagonal lattice parameter $d_{hexagonal}(T)$ (Eq. 23) for the $MgB_2$ film.

The $\left|\overrightarrow{F_p}(B,T)\right|$ data fits to Eq. 3 and deduced parameters are showed in Figures 5 and 6 respectively. It can be seen (Fig. 6) that both deduced parameters, $p(T)$ and $q(T)$, are varying within very wide ranges which were not described by Dew-Hughes [5]. Moreover, $p(T)$ and of $q(T)$ parameters deduced at different temperatures indicate different pinning mechanisms. For instance, $p(T = 25\ K) = 0.36$ and $q(T = 25\ K) = 1.1$ which imply that the pinning is due to magnetic/volume/normal mechanism [5], while $p(T = 10\ K) = 0.4$ and $q(T = 10\ K) = 1.9$ which imply that the pinning is due to core/surface/normal mechanism [5]. This result demonstrates that the Kramer-Dew-Hughes [4,5] model exhibits a problem with the validity of deduced values interpretation, while mathematical fits to Eq. 3 are, as a rule, reasonably accurate approximated $\left|\overrightarrow{F_p}(B)\right|$ data.

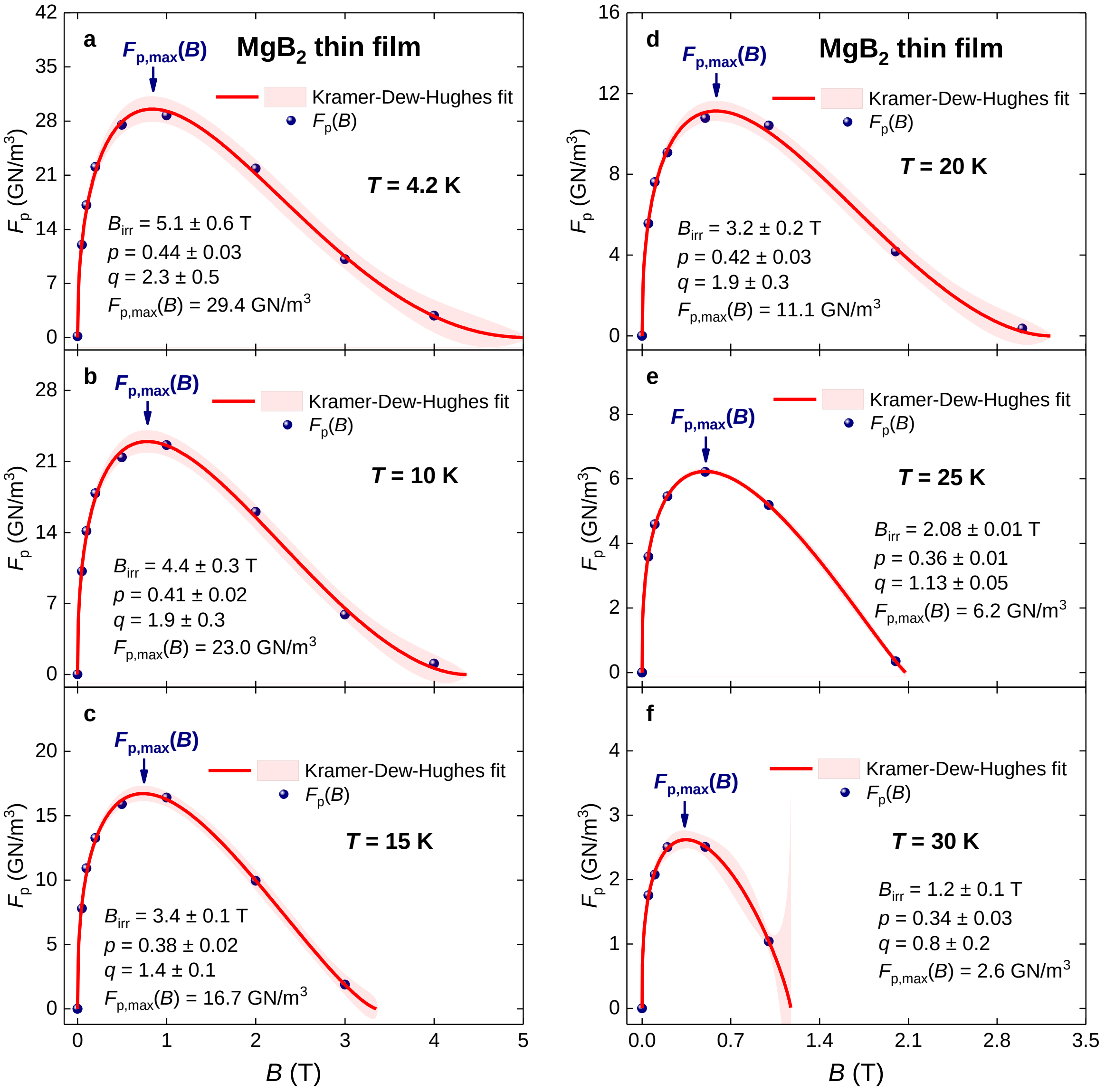

**Figure 5.** $\left|\overrightarrow{F_p}(B, T, \theta = 0°)\right|$ curves and data fits to Eq. 13 for $MgB_2$ thin film deposited on a 6H-SiC single crystal substrate for temperature range 4.2-30 K (a-f). Raw transport $J_c(B, T, \theta = 0°)$ data was reported by Zheng *et al* [21]. Goodness of fit for all panels is better than $R$ = 0.995. 95% confidence bands are shown by shadow areas.

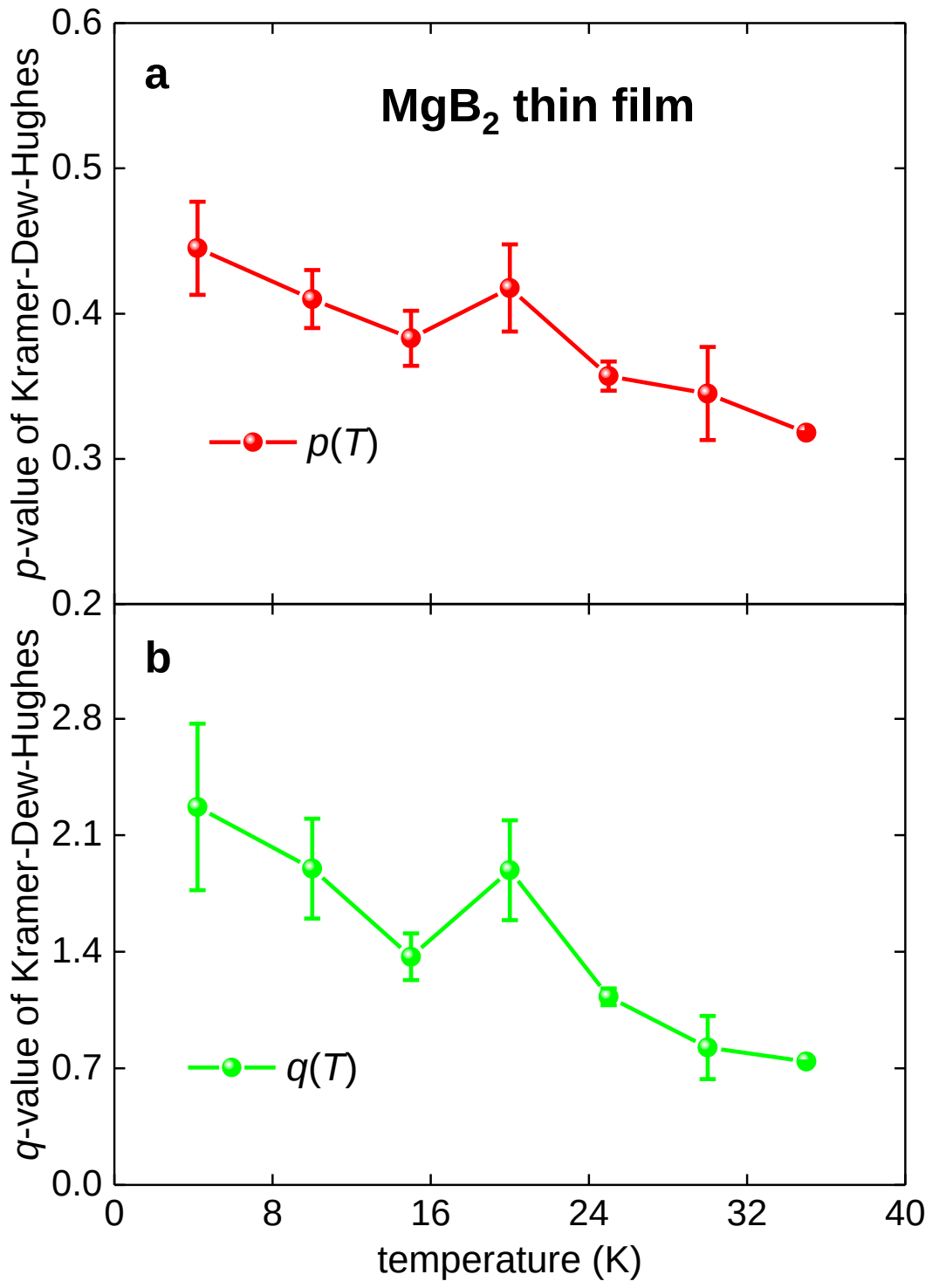

**Figure 6.** Deduced free-fitting parameter *p*(*T*) (a) and *q*(*T*) (b) of the Kramer-Dew-Hughes model (Eq. 5) for $MgB_2$ thin film deposited on a 6H-SiC single crystal substrate for which raw transport $J_c(B,T,\theta = 0°)$ data was reported by Zheng *et al* [21].

### 3.2. Pnictides thin films

In this Section, the analysis was applied for two NdFeAs(O,F) thin films for which raw transport critical current density data, $J_c(B,T,\theta = 0°)$, reported by Tarantini *et al* [41] and by Guo *et al* [42]. Both research groups found that the $\left|\overrightarrow{F_p}(B,T,\theta = 0°)\right|$ data fit to Eq. 3 generates unphysical $p(T)$ and $q(T)$ values, which was interpreted as a demonstration of a superposition of two different pinning mechanisms. To resolve this problem Tarantini *et al* [41] proposed to fit $\left|\overrightarrow{F_p}(B,T,\theta = 0°)\right|$ datasets to a function which is a sum of two Kramer-Dew-Hughes functions, where one function accumulates the contribution of the surface pinning (for which $p(T) \equiv 0.5$ and $q(T) \equiv 2.0$ in accordance with Ref. 5) and another function accumulates the contribution of the point pining (for which $p(T) \equiv 1.0$ and $q(T) \equiv 2.0$ in accordance with Ref. 5):

$$\left|\overrightarrow{F_p}(B)\right| = F_{p,max,s} \times \left(\frac{B}{B_{irr,s}}\right)^{0.5} \times \left(1 - \frac{B}{B_{irr,s}}\right)^{2} + F_{p,max,pd} \times \left(\frac{B}{B_{irr,pd}}\right)^{1.0} \times \left(1 - \frac{B}{B_{irr,pd}}\right)^{2} \quad (26)$$

Slightly different approach was proposed by Guo *et al* [42], who considered that two summation parts in Eq. 26 have $q(T) \equiv 2.0$ and, thus, these authors proposed to use the following fitting function:

$$\left|\overrightarrow{F_p}(B)\right| = F_{p,max} \times \left(\frac{B}{B_{irr}}\right)^p \times \left(1 - \frac{B}{B_{irr}}\right)^2 \tag{27}$$

where $p$ is free fitting parameter. Because the surface pinning characterizes by $p \equiv 0.5$ and the point pining characterizes by $p \equiv 1.0$ [5], then the proximity of free-fitted $p$ value to either 0.5 or 1.0 can be interpreted as the dominance of the surface pinning or the point pinning.

In the results, both research groups [41,42] reported that the point pinning dominates at $0.5 \lesssim \frac{T}{T_c}$, while at $\frac{T}{T_c} \lesssim 0.5$ the dominant mechanism is the surface pining. Because the comprehensive analysis of $\left|\overrightarrow{F_p}(B, T, \theta = 0°)\right|$ for NdFeAs(O,F) thin films was reported in Refs. 41,42 here we only presented the analysis for the $\left|\overrightarrow{F_p}(J_c, T, \theta = 0°)\right|$ datasets.

In Figure 7 we showed $\left|\overrightarrow{F_p}(J_c, B, T = 35\ K, \theta = 0°)\right|$ dataset for NdFeAs(O,F) thin film for which $J_c(B, T, \theta = 0°)$ was reported by Guo *et al* [42]. Three orthogonal projections of the $\left|\overrightarrow{F_p}(J_c, B, T = 35\ K, \theta = 0°)\right|$ curve are shown in Figure 8. $\left|\overrightarrow{F_p}(B)\right|$ curve (green) can be fitted to Eq. 26 with a high quality (Fig. 8). However, on this projection an important property of $\left|\overrightarrow{F_p}(J_c, B, T = 35\ K)\right|$ dataset, which is a linear dependence of $\left|\overrightarrow{F_p}(J_c, B, T = 35\ K)\right|$ vs $J_c$ (Fig. 7 and Fig. 8(c)), is invisible.

Another important feature, which makes $\left|\overrightarrow{F_p}(J_c, B, T)\right|$ curve for NdFeAs(O,F) thin film different from its counterpart for $MgB_2$ film (Figs. 1,2) is that the low-field linear part of $\left|\overrightarrow{F_p}(J_c, B, T)\right|$ curve abruptly transforms into a dome-like shape of $\left|\overrightarrow{F_p}(J_c, B, T)\right|$ at $J_{c,Fp,switch}$ point. This important feature is also invisible in the $\left|\overrightarrow{F_p}(B)\right|$ projection of the $\left|\overrightarrow{F_p}(J_c, B, T)\right|$ curve (Fig. 8(b)).

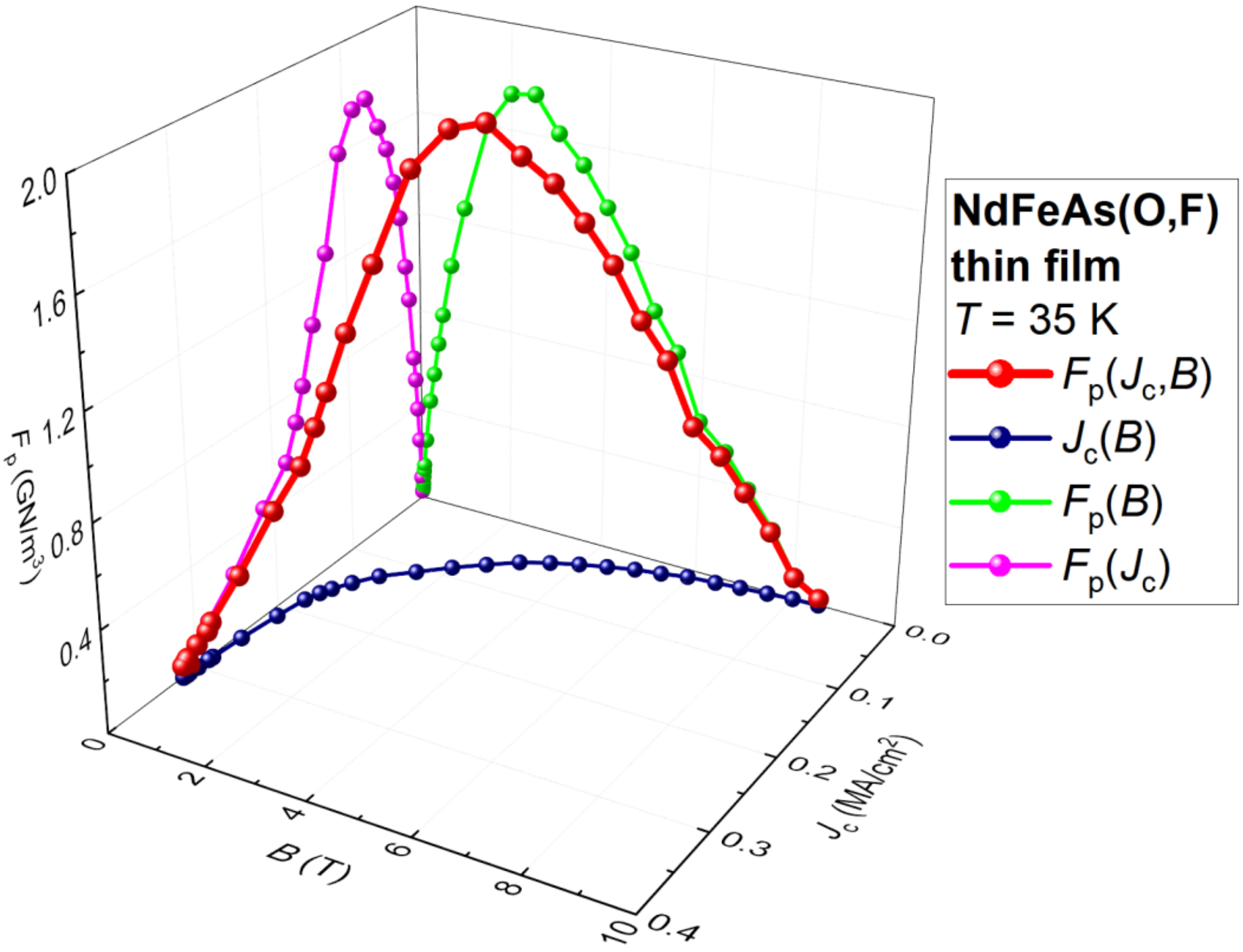

**Figure 7.** Three-dimensional (3D) representation of the pinning force density, $\left|\overrightarrow{F_p}(J_c, B, T = 35\ K, \theta = 0°)\right|$, NdFeAs(O,F) thin film for which $J_c(B, T, \theta = 0°)$ was reported by Guo *et al* [42].

One can see in Figs. 7,8 that 3D $\left|\overrightarrow{F_p}(J_c, B, T)\right|$ curve starts to deviate from the $\left|\overrightarrow{F_p}(J_c, T)\right|$ projection when $J_c \leq J_{c,Fp,switch}$. For $J_c \leq J_{c,Fp,max}$ the 3D $\left|\overrightarrow{F_p}(J_c, B, T)\right|$ curve is better approximated by $\left|\overrightarrow{F_p}(B, T)\right|$ projection. The fitting function for the dome-like part of the $\left|\overrightarrow{F_p}(J_c, T)\right|$ curve should be flexible within the full range of $J_c \leq J_{c,Fp,switch}$. We constructed several possible functions and found that the general form proposed by Kramer [4] and by Dew-Hughes [5] can reasonably well fit $\left|\overrightarrow{F_p}(J_c, T)\right|$ data:

$$F_p(J_c) = \theta\left(J_c \leq J_{c,Fp,switch}\right)\left(\frac{J_c}{J_{c,1}}\right)^l\left(1 - \left(\frac{J_c}{J_{c,1}}\right)\right)^m + \theta\left(J_c \geq J_{c,Fp,switch}\right)(c_2 \times J_c + f_2) \quad (28)$$

where $J_{c,Fp,switch}$, $J_{c,1}$, $l$, $m$, $c_2$ and $f_2$ are free-fitting parameters. It should be stressed, that because (as one can see in Figs. 1,7) the projection of the 3D $\left|\overrightarrow{F_p}(J_c, B, T)\right|$ curve into the $B =$ 0 plane is significantly distorted from its original 3D curve, when $0 \leq J_c \leq J_{c,Fp,max}$. And the

fit to Eq. 28 can be a good approximation for $J_c$ values within the range of $J_{c,Fp,max} \leq J_c \leq J_c(sf)$.

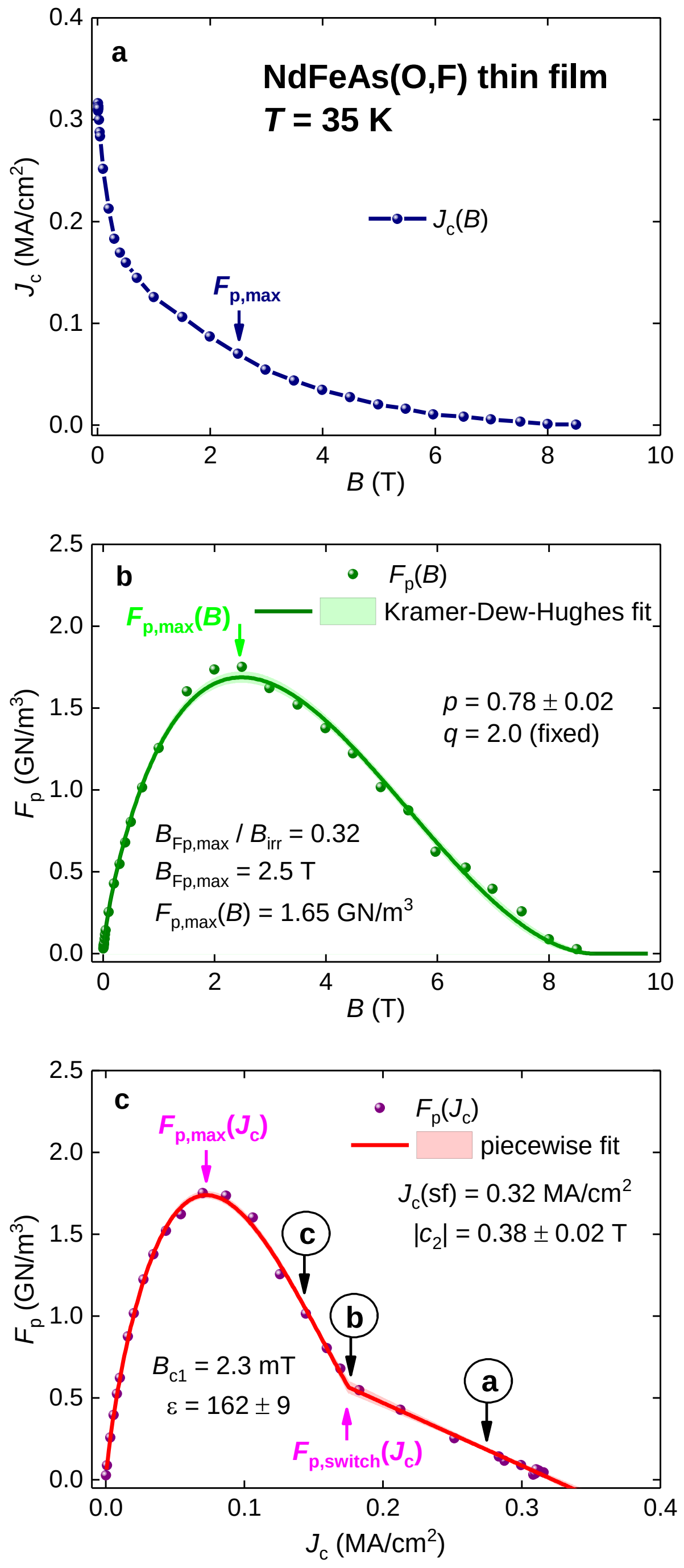

**Figure 8.** Projections of the $\left|\overrightarrow{F_p}(J_c, B, T = 35\ K, \theta = 0°)\right|$ curve for NdFeAs(O,F) thin film into three orthogonal planes. Data fit to Eq. 27 (panel **b**) and Eq. 30 (panel **c**) are shown. **a** – the projection into the $F_p$=0 plane; **b** – the projection into the $J_c$=0 plane and data fit to Kramer-Dew-Hughes (equation 3); and **c** – the projection into the $B$=0 plane and data fit to piecewise model (equation 28). Deduced parameters are: $\frac{J_{c,Fp,max}}{J_c(sf)} = 0.23$, $\frac{J_{c,Fp,switch}}{J_c(sf)} = 0.53$, $J_{c,Fp,max} = 0.07\ \frac{MA}{cm^2}$, $J_{c,Fp,switch} = 0.18\ \frac{MA}{cm^2}$, $J_c(sf) = 0.32\ \frac{MA}{cm^2}$. Numbers 1,2,3 in panel ***c*** respect to the schematic representation of the vortex lattice structure showed in Fig. 9. Raw $J_c(B, T = 35\ K, \theta = 0°)$ data was reported by Guo *et al* [42]. 95% confidence bands are shown by shadow areas.

Thus, there is an interesting analogy with the Kramer-Dew-Hughes fit (Eq. 3), which is a reasonably accurate approximation for $\left|\overrightarrow{F_p}(J_c, B, T)\right|$ curve for large applied magnetic fields, i.e., within a range of $B_{Fp,max} \leq B \leq B_{c2}$, while Eq. 28 is a good approximation for $\left|\overrightarrow{F_p}(J_c, B, T)\right|$ for large critical current densities, $J_{c,Fp,max} \leq J_c \leq J_c(sf)$, or what is the equivalent statement, for low- and middle-range applied magnetic fields $0 \leq B \leq B_{Fp,max}$.

However, we can note that our Eq. 28 has a linear part, which represents a significant difference between Eq. 3 and Eq. 28. In addition, as we stated above, there is a need to revise the validity of the primary interpretation of *p* and *q* parameters in Kramer-Dew-Hughes fit, which was proposed Dew-Hughes nearly five decades ago [5]. Based on this concern, we do not provide any interpretation for *l* and *m* parameters in Eq. 28 now.

To interpret the physical origin of the appearance of the dome-like bump in the $\left|\overrightarrow{F_p}(J_c, T)\right|$ datasets (Figs. 9), which is suddenly appeared at $J_c \leq J_{c,Fp,switch}$ (Fig. 8(c)), we can propose that this dome is due to the effect of the Abrikosov's vortices collective pinning [43]. This effect was theoretically predicted by Larkin and Ovchinnikov more than five decades ago [43], and despite a wide discussion of this effect in cuprates [44], from the author's best knowledge Fig. 8 represents clear experimental evidence for the existence of this effect. This interpretation is well aligned with the absence of the collective pinning in $MgB_2$ and, thus, Figs. 1-3 demonstrate that the dome-like bump in the $\left|\overrightarrow{F_p}(J_c, T)\right|$ datasets does not exist.

Following general idea for the magnetic flux distribution in thin superconducting slab when external magnetic field is applied in perpendicular direction to the film surface proposed by Brandt and Indenbom [45], in Figure 9 we showed schematic representation for vortices lattice for three distinguish part of the $\left|\overrightarrow{F_p}(J_c)\right|$ curve which can be seen in Fig. 8(c). At low applied magnetic field, i.e. $B < |c_2|$, Abrikosov's vortices fill the superconducting film starting from the slab edges and vortices is separated by the distance, $d_{hexagonal}$, which

corresponds to the $|c_2|$ field (Eq. 23). This stage corresponds to the linear part of the the $|\overrightarrow{F_p}(J_c)|$ curve indicated by number "1" in Fig. 8(c).

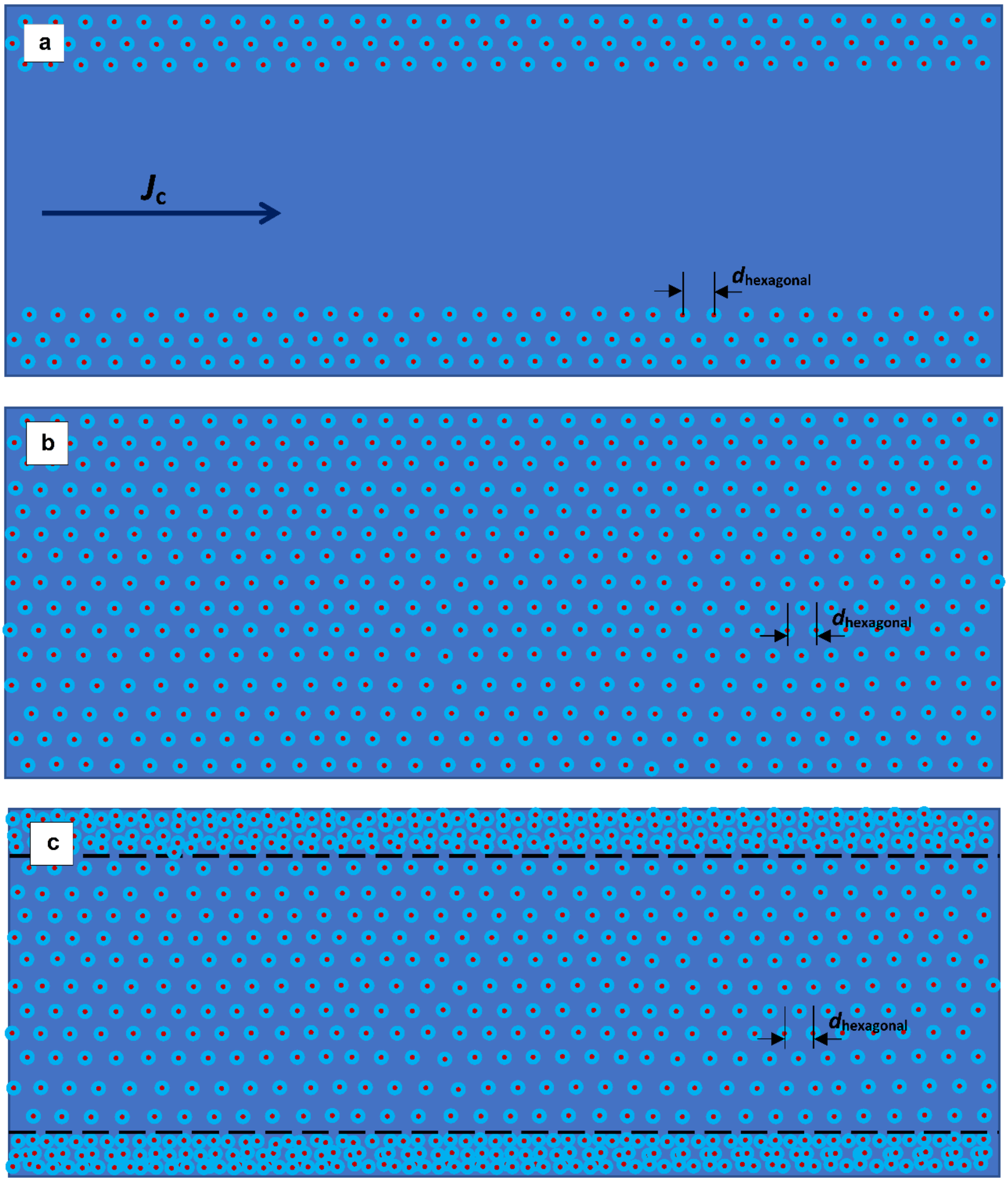

**Figure 9.** Schematic representation of the vortex structure when magnetic field is applied in the perpendicular direction to the thin film surface. **a** – the vortex structure at low applied magnetic field, $B < |c_2|$, which corresponds to the linear part (marked as "1" in Figure 8(c)) of the $|\overrightarrow{F_p}(J_c)|$ pinning force density. The distance between vortices, $d_{hexagonal}$, corresponds to the $|c_2|$ field (Eq. 23). **b** – the vortex structure at applied field $B \cong B_{switch} \cong |c_2|$ (this stage is depicted by arrows "2" and $F_{p,switch}(J_c)$ in Figure 8(c). **c** – the vortex structure at applied field $B > |c_2|$ when the material exhibits collective flux pining effect. This stage schematically showed by arrow "3" in Figure 8(c).

The vortices completely and uniformly fill the film at applied field of $B \cong B_{switch} \cong |c_2|$, while the distance between vortices keeps to be $d_{hexagonal}$ (Eq. 23). This stage is indicated by number "2" in Figure 8 and the schematic diagram is shown in Fig. 9(b).

At applied $B > |c_2|$ there are two possible scenarios. One is for materials like $MgB_2$ which do not exhibit the collective flux pinning effect, and after the applied field reaches $|c_2|$ the pinning force density is rapidly dropped (see for instance Fig. 3). Otherwise, if material exhibits the collective flux pinning effect, Abrikosov's vortices penetrate in the slab from its edges and form the second flux penetration front, which has much higher fluxons density and does not exhibit hexagonal flux lattice structure.

The fit of the $\left|\overrightarrow{F_p}(J_c, T = 35\ K)\right|$ dataset to Eq. 28 and calculations based on Eqs. 18,21,23 reveal the following values (for calculations we used the Ginsburg-Landau parameter $\kappa(T) = 90$ [46]):

$$|c_2(T = 35\ K)| = 0.38 \pm 0.02\ T \quad (29)$$

$$B_{c1}(T = 35\ K) = 2.3\ mT \quad (30)$$

$$\varepsilon(T = 35\ K) = 16.5 \quad (31)$$

$$d_{hexagonal}(T = 35\ K) = 79\ nm \quad (32)$$

Fits to Eq. 28 of $\left|\overrightarrow{F_p}(J_c, T)\right|$ datasets measured at different temperature for the NdFeAs(O,F) thin film are shown in Figs. 10,11. Overall, $\left|\overrightarrow{F_p}(J_c, T)\right|$ datasets measured at high reduced temperatures, $\frac{T}{T_c}$, have a very pronounced linear part (Fig. 11). When the datasets do not contain large enough raw data points to be fitted to Eq. 28 (for all parameters are free), we fixed $m = 2.0$ value adopting the approach proposed for NdFeAs(O,F) thin films in Refs. 36,37. The primary reason for this is that our purpose is to extract as accurately as possible the $|c_2(T)|$ value, which is the $F_p(J_c)$ slope at a low applied magnetic field. Based on this, the accuracy of the approximation of the collective pinning peak in the $F_p(J_c)$ cannot be characterized as our primary task, because we target to deduce the values listed in Eqs. 29-32, for which parameters describing the collective pinning bump are not required.

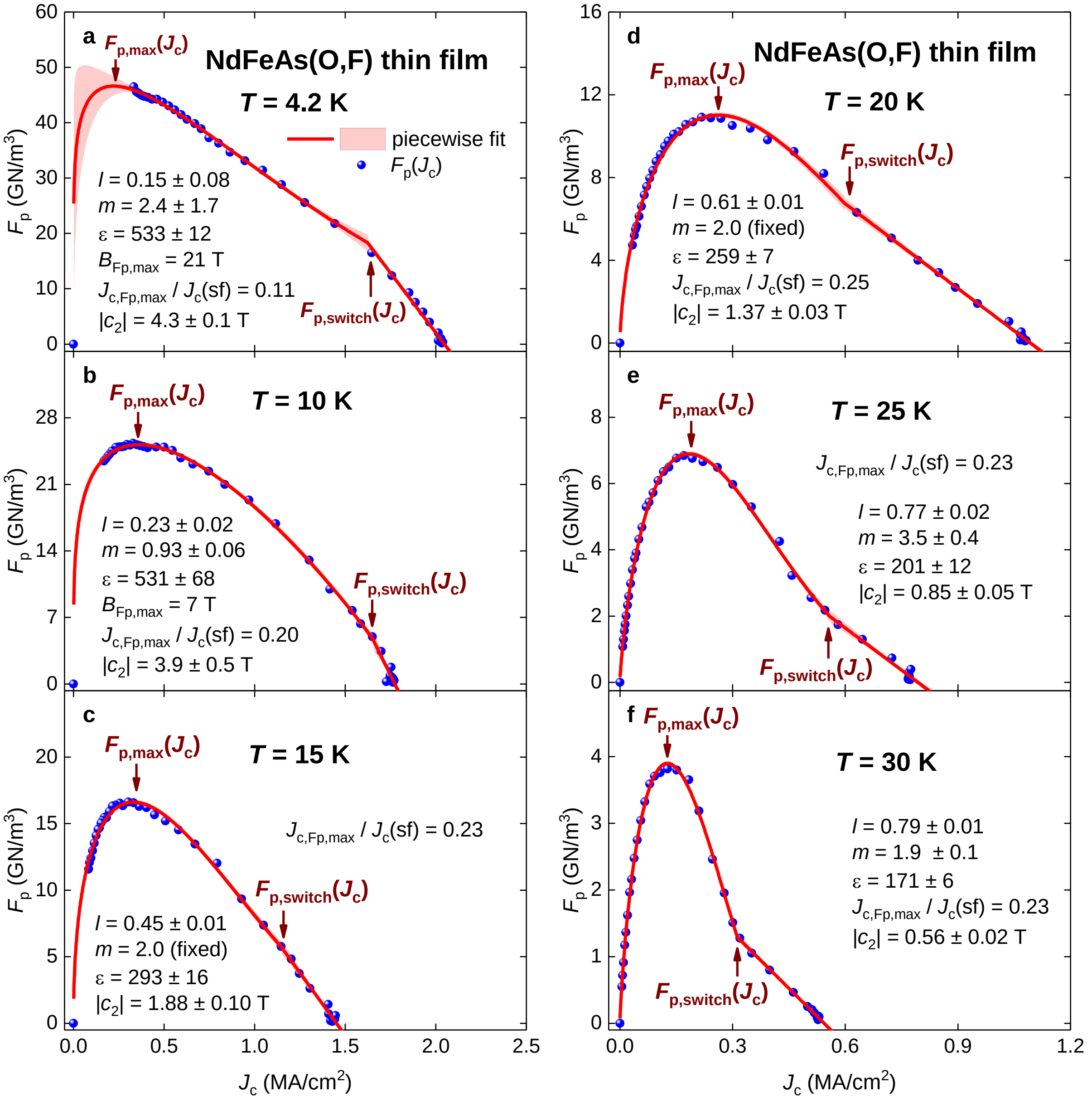

**Figure 10.** $\left|\overrightarrow{F_p}(J_c,T,\theta=0°)\right|$ curves and data fits to Eq. 28 for NdFeAs(O,F) thin film measured at temperature range 4.2-30 K (a-f). Raw transport $J_c(B,T,\theta=0°)$ data was reported by Guo *et al* [42]. Goodness of fit for all panels is better than $R$ = 0.998. 95% confidence bands are shown by shadow areas.

It is important to note that all fits to Eq. 28 of all reported (by Guo *et al* [42]) $F_p(J_c)$ datasets were converged which is a prominent advantage versus $F_p(B)$ datasets fit to Eq. 5, where the converging is the main issue, especially for datasets measured at low temperatures.

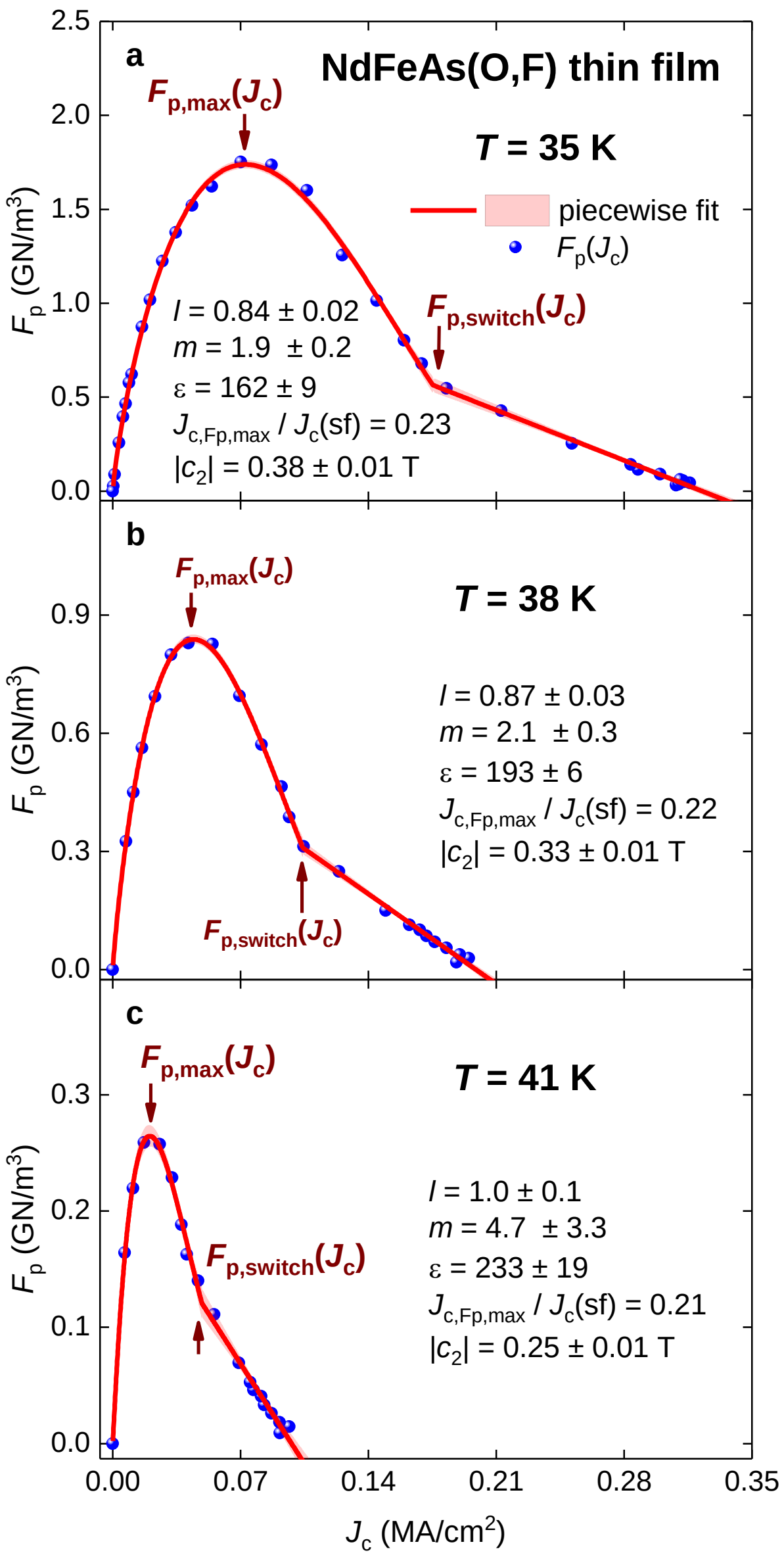

**Figure 11.** $|\overrightarrow{F_p}(J_c, T, \theta = 0°)|$ curves and data fits to Eq. 28 for NdFeAs(O,F) thin film measured at temperature range 35-41 K. Raw transport $J_c(B, T, \theta = 0°)$ data was reported by Guo *et al* [42]. Goodness of fit for all panels is better than $R$ = 0.998. 95% confidence bands are shown by shadow areas.

The converged $F_p(J_c)$ curve can be inverted to calculate the important pinning parameter $B_{Fp,max}$, which can not be extracted from $F_p(B)$ datasets, because the latter dependences, as a rule, at $T$ ~ 4 K for all practically important materials (i.e., cuprates and IBS) are increasing function of $B$. Based on this, the $F_p(B, T = 4.2\ K)$ fits to Eq. 3 diverge.

However, if the same $|\overrightarrow{F_p}(J_c, B, T = 4.2\ K)|$ dataset is to be projected into the $B$ = 0 plane (which is $|\overrightarrow{F_p}(J_c, T = 4.2\ K)|$), it can converge, as is demonstrated in Fig. 10(a). Because the $\left|\overrightarrow{F_p}(J_c, T = 4.2\ K)\right|$ curve, as we mentioned above, is an accurate approximation for the 3D $\left|\overrightarrow{F_p}(J_c, B, T = 4.2\ K)\right|$ curve for $0 \le B \le B_{Fp,max}$, the deduced $B_{Fp,max}(T = 4.2\ K)$ from converged fit to Eq. 28 can be a reasonably accurate estimation for $B_{Fp,max}(T = 4.2\ K)$. To perform this inversion of the fitted $\left|\overrightarrow{F_p}(J_c, T = 4.2\ K)\right|$ curve for NdFeAs(O,F) film (Fig. 10(a)), one obtained $B_{Fp,max}(T = 4.2\ K) = 21\ T$, which is significantly above the highest applied magnetic field in the given experiment, $B = 14\ T$ [42].

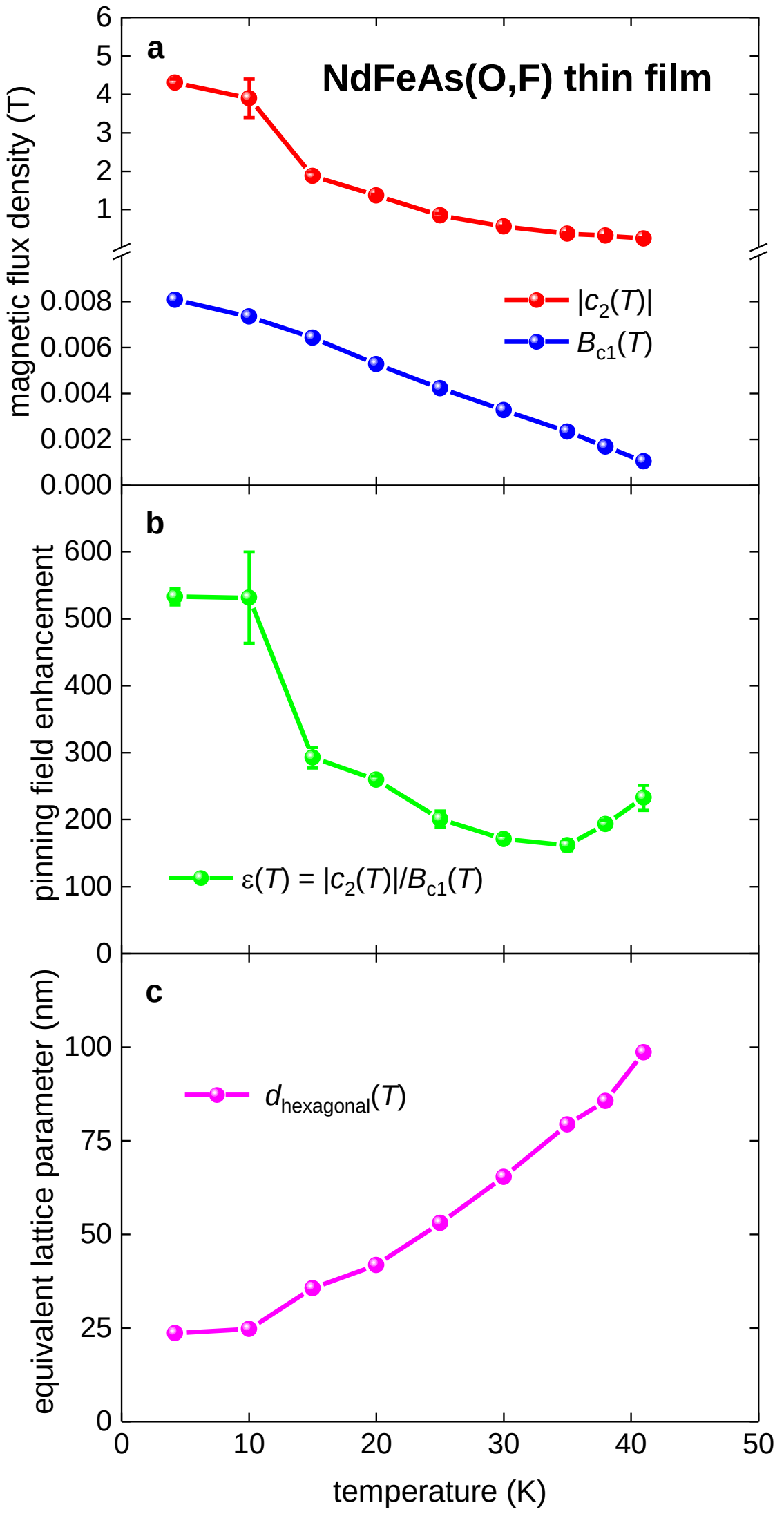

**Figure 12. a** - Deduced free-fitting parameter $|c_2(T)|$ (Eq. 13) and $B_{c1}(T)$ calculated by Eq. 21 for NdFeAs(O,F) thin film for which raw transport $J_c(B, T, \theta = 0°)$ data was reported by Guo *et al* [38]. **b** – The pining field enhancement factor $\varepsilon(T)$ (Eq. 18) for the same film. **c** – Equivalent hexagonal lattice parameter $d_{hexagonal}(T)$ (Eq. 23) for the same film.

Primary deduced parameters are shown in Figure 12. There are remarkable high values for the pinning field enhancement factor, $\varepsilon(T)$, within full temperature range. This is the most distinguish difference in values deduced for NdFeAs(O,F) thin film in comparison with $MgB_2$ (see for instance, Figs. 4,12).

### 3.3. REBCO thin films

The introduction of nano-scaled secondary phases (so-called artificial pinning centres (APC)) in REBCO thin films is conventional approach to enhance the in-field critical current density, $J_c(B,T)$, which was introduced nearly simultaneously by MacMagnus-Driscoll *et al* [47] and Haugan *et al* [48]. However, there are growing number of experimental evidences [49-50] that prominent $J_c(B,T)$ enhancement exhibits only at high reduced temperature range, i.e., at $0.7 \lesssim \frac{T}{T_c}$, while at $\frac{T}{T_c} \lesssim 0.25$ the enhancement originated from APC and 124-stacking faults practically vanish [49,50]. In attempts to develop better understanding of this problem, which has a great practical importance [51], we analysed $\left|\overrightarrow{F_p}(J_c,B,T)\right|$ data for REBCO thin films by the approach described above.

In Figure 13 we showed 3D $\left|\overrightarrow{F_p}(J_c,B,T=75.5\ K)\right|$ data curve for undoped $YBa_2Cu_3O_{7-x}$ thin film (Sample 87) reported by MacMagnus-Driscoll *et al* [47] (it should be noted that in Figure 1(b) [47] (from which $J_c(B,T=75.5\ K)$ data was digitized) the Y-axis has a mistake in value numbering).

Three orthogonal projections of $\left|\overrightarrow{F_p}(J_c,B,T=75.5\ K)\right|$ curve are shown in Figure 13(b), where one can see that the fit to Eq. 3 of $\left|\overrightarrow{F_p}(B)\right|$ dataset has a high quality. However, a linear dependence of $\left|\overrightarrow{F_p}(J_c,B,T=75.5\ K)\right|$ vs $J_c$ (Figs 13, 14(c,d)), which covers significant part of $0.25 \lesssim \frac{J}{J_c(sf)} \lesssim 1.0$ range, cannot be recognized in the $\left|\overrightarrow{F_p}(B)\right|$ projection.

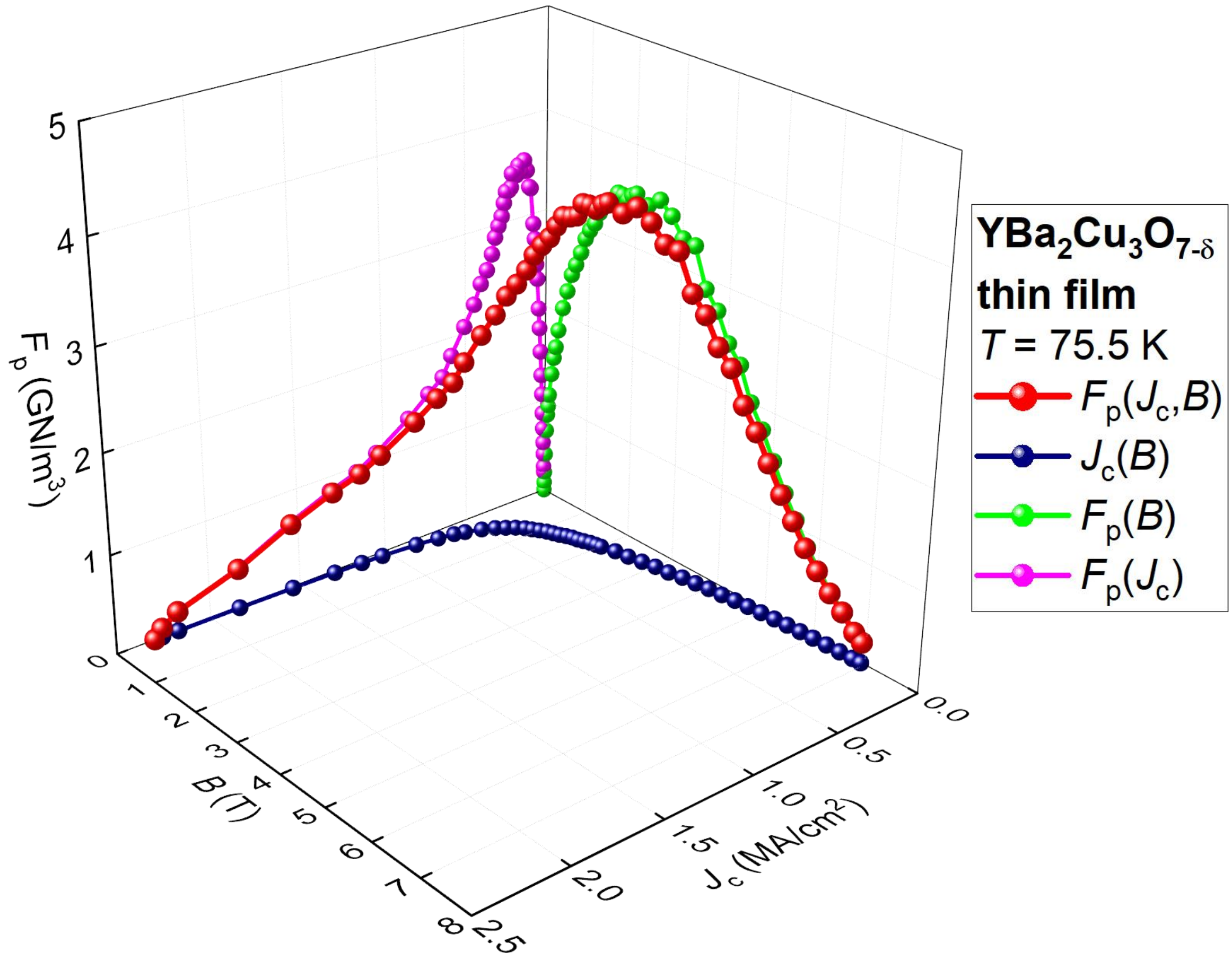

**Figure 13.** Three-dimensional (3D) representation of the pinning force density, $\left|\overrightarrow{F_p}(J_c, B, T = 75.5\ K, \theta = 0°)\right|$, for undoped $YBa_2Cu_3O_{7-\delta}$ thin film (Sample 87 [47]). Raw $J_c(B, T = 75\ K, \theta = 0°)$ data was reported by MacMagnus-Driscoll *et al* [47].

The fit of $\left|\overrightarrow{F_p}(J_c, T = 75.5\ K)\right|$ dataset to Eq. 28 is shown in Fig. 14(c), where $m$ parameter was fixed to $m = 2.0$. However, a more accurate approximation of the linear part and the full $\left|\overrightarrow{F_p}(J_c, T = 75.5\ K)\right|$ curve is achieved by using a three-step linear piecewise function:

$$F_p(J_c) = \theta(J_c \leq J_{c,Fp,max}) \times (c_0 \times J_c + f_0) + \theta(J_c \geq J_{c,Fp,max}) \times \theta(J_c \leq J_{c,Fp,switch}) \times (c_1 \times J_c + f_1) + \theta(J_c \geq J_{c,Fp,switch}) \times (c_2 \times J_c + f_2). \quad (33)$$

The fit is shown in Fig. 14(d).

Eq. 33 represents one of the simplest and, simultaneously, remarkably accurate mathematical expression of our primary idea that $\left|\overrightarrow{F_p}(J_c, B, T)\right|$ has three distinctive ranges (which we already mentioned above, but it is important listed these ranges again), each of those can be approximate by a linear curve:

1. low reduced applied magnetic field range, where the approximation is $|\overrightarrow{F_p}(J_c)|$ line (which corresponds to the schematic diagram showed in Fig. 9(a));
2. high reduced applied magnetic field range, where the approximation is $|\overrightarrow{F_p}(B)|$ curve;
3. middle reduced applied magnetic field, where $|\overrightarrow{F_p}(J_c)|$ line and $|\overrightarrow{F_p}(B)|$ curve can be approximated by a median curve or line.

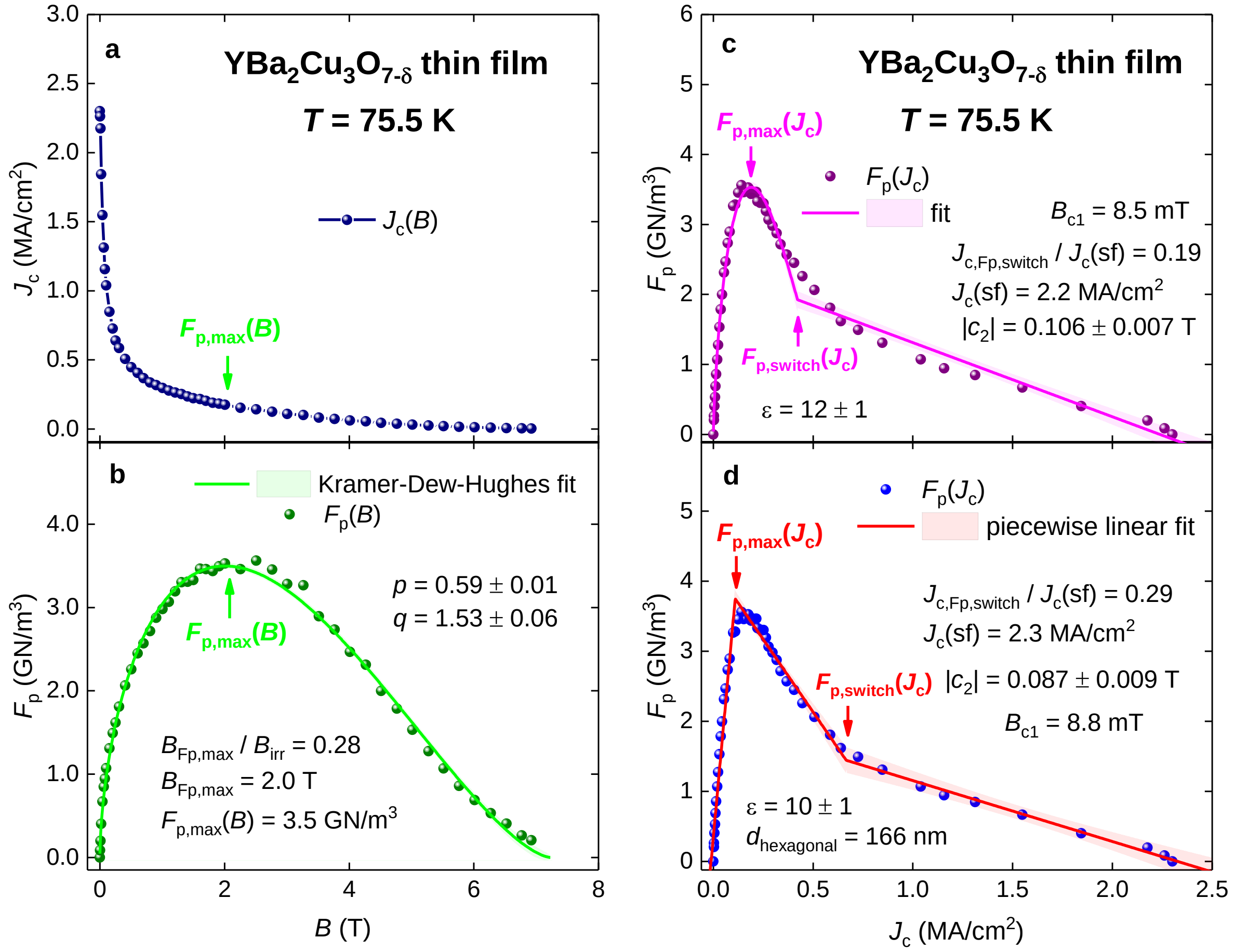

**Figure 14.** Projections of the $|\overrightarrow{F_p}(J_c, B, T = 75.5\ K, \theta = 0°)|$ curve for $YBa_2Cu_3O_{7-\delta}$ thin film into three orthogonal planes. Data fit to Eq. 3 (panel **b**), Eq. 28 (panel **c**), and Eq. 33 (panel **d**) are shown. **a** – the projection into the $F_p$=0 plane; **b** – the projection into the $J_c$=0; **c** and **d** – the projection into the $B$=0 plane. For panel **c** parameters in Eq. 28 are: $l = 0.73 \pm 0.2$, and $m = 2.0$ (fixed). The goodness of fit is $R$ = 0.988. For panel **d**, the goodness of fit is $R$ = 0.985. Raw $J_c(B, T = 75.5\ K, \theta = 0°)$ data was reported by MacMagnus-Driscoll *et al* [47]. 95% confidence bands are shown by shadow areas.

Calculations based on Eqs. 21,23,33 reveal the following values (for calculations we used the Ginsburg-Landau $\kappa(T) = 95$ [52]):

$$|c_2(T = 75.5\ K)| = 0.087 \pm 0.009\ T \quad (34)$$

$$B_{c1}(T = 75.5\ K) = 8.8\ mT \quad (35)$$

$$\varepsilon(T = 75.5\ K) = 10 \pm 1 \quad (36)$$

$$d_{hexagonal}(T = 77.5\ K) = 166\ nm \quad (37)$$

Before we present results of analysis of temperature dependence of the $\left|\overrightarrow{F_p}(J_c, T)\right|$ datasets for one of a few commercially available HTS 2G-wires, it will be interesting to present the $\left|\overrightarrow{F_p}(J_c, T = 4.2\ K)\right|$ data analysis for laboratory made sample, which can be considered as the upper limiting case for commercial HTS 2G-wires. Xu *et al* [53] fabricated and studied ~0.9 μm thick (Gd,Y)$Ba_2Cu_3O_{7-\delta}$ + 15% $BaZrO_3$ thin film which was deposited on a standard buffered IBAD Hastelloy substrate by metal-organic chemical vapor deposition technique. The $J_c(B, T = 4.2\ K)$ and the $\left|\overrightarrow{F_p}(B, T = 4.2\ K)\right|$ datasets are shown in Figure 14(a,b) respectively (datasets were reported in the Ref. 53). It should mention that the $\left|\overrightarrow{F_p}(B, T = 4.2\ K)\right|$ fit to the Kramer-Dew-Hughes model, despite it converged at fixed $q$ =2.0 value, does not have high quality.

However, the fit of $\left|\overrightarrow{F_p}(J_c, T = 4.2\ K)\right|$ dataset to Eq. 33 has converged (Fig. 14(c)), and the linear dependence of $\left|\overrightarrow{F_p}(J_c, T = 4.2\ K)\right|$ at high critical current densities is remarkably clear and accurate. It should be stressed, that there is no visual designations either in Fig. 15(a), nor Fig. 15(b) that there is a very sharp transition between two linear dependences in $\left|\overrightarrow{F_p}\right|$ vs $J_c$, which can be clear observed in Fig. 14(c) at $J_c(T = 4.2\ K) = 23\ \ MA/cm^2$.

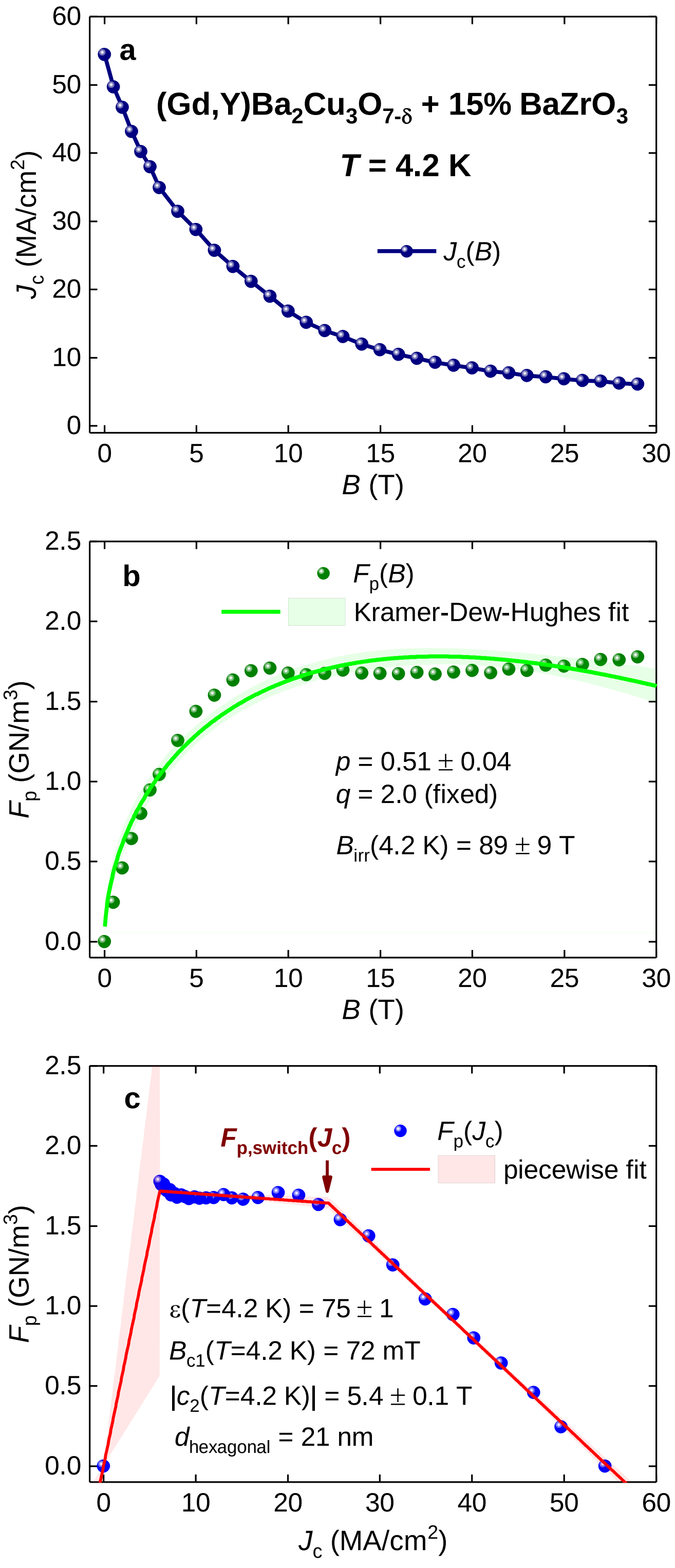

**Figure 15.** Projections of the $\left|\overrightarrow{F_p}(J_c, B, T = 4.2\ K, \theta = 0°)\right|$ curve for (Gd,Y)$Ba_2Cu_3O_{7-\delta}$ + 15% $BaZrO_3$ thin film into three orthogonal planes. Data fit to Eq. 3 (panel **b**) and Eq. 33 (panel **c**) are shown. Raw $J_c(B, T = 4.2\ K)$ data was reported by Xu *et al* [53]. Goodness of fit (b) $R$ = 0.95 and (c) $R$ = 0.9975. 95% confidence bands are shown by shadow areas.

Calculations based on Eqs. 21,23 reveal the following values (for calculations we used the Ginsburg-Landau $\kappa(T) = 95$ [52]):

$$|c_2(T = 4.2\ K)| = 5.4 \pm 0.1\ T \tag{38}$$

$$B_{c1}(T = 4.2\ K) = 72\ mT \tag{39}$$

$$\varepsilon(T = 4.2\ K) = 75 \pm 1 \tag{40}$$

$$d_{hexagonal}(T = 4.2\ K) = 21\ nm \tag{41}$$

There are two good matches between deduced values (Eqs. 38-41) and reported values. The first one is deduced $B_{c1}(T = 4.2\ K) = 72\ mT$, which is in a good agreement with reported values for $YBa_2Cu_3O_{7-\delta}$ single crystals [54,55].

The second agreement is with the reported by Xu *et al* [53] the average distance between $BaZrO_3$ nanorods in REBCO matrix which was extracted from the analysis of transmission electron microscopy (TEM) images, $d_{average,TEM} \sim 17 - 25\ nm$ [53]. We deduced the equivalent hexagonal vortex lattice parameter $d_{hexagonal}(T = 4.2\ K) = 21\ nm$ (Fig. 15 and Eqs. 23,33). These two characteristic lengths can be considered be equal, because of natural variation of the $BaZrO_3$ nanorods density in REBCO matrix has reasonable variation, even within the viewing area of the same TEM image (see, for instance, Fig. 10 in Ref. 15).

Thus, there is the first direct experimental evidence that the $|c_2(T)|$ field can be interpreted as the matching field related to the density of structural defects in superconductors.

We should stress, that the $|c_2(T)|$ field is not a field at which the $\left|\overrightarrow{F_p}(J_c, B, T)\right|$ or the $J_c(B,T)$ curves have any sort of inflection or other unique features, because the $|c_2(T)|$ field is the linear coefficient between the $\left|\overrightarrow{F_p}(J_c, B, T)\right|$ and $J_c$ at low reduced applied magnetic field, $\frac{B}{B_{irr}}$. Perhaps this is the primary point which designates our approach and previous attempts to define the matching field. This means, that the $|c_2(T)|$ is the characteristic field of the superconductor within a wide range of applied magnetic field, $B$, and there are no any designated features in the $\left|\overrightarrow{F_p}(J_c, B, T)\right|$ and the $J_c(B,T)$ curves at applied magnetic field of $B = |c_2(T)|$. This is obvious, but important issue, that the $|c_2(T)|$ is not a field at which the inflection in $\left|\overrightarrow{F_p}(J_c, T)\right|$ curve (i.e., $\left|F_{p,switch}(J_c, T)\right|$) is observed. From the other hand, it might be useful to perform detailed experimental studies of the $\left|\overrightarrow{F_p}(J_c, B \cong |c_2(T)|, T)\right|$ and the $|J_c(B \cong |c_2(T)|, T)|$ in a variety of superconductors.

It is interesting to note that deduced $d_{hexagonal}(T = 4.2\ K) = 21\ nm$ (Eq. 41) for (Gd,Y)$Ba_2Cu_3O_{7-\delta}$+15% $BaZrO_3$ film is in the same ballpark with $d_{hexagonal}(T = 4.2\ K) = 24\ nm$ for NdFeAs(O,F) film which we deduced above (Fig. 12).

To demonstrate of the applicability of our approach to analyse commercial HTS 2G-wires, in Figures 16,17 we showed the analysis for SuperPower Inc. wire (raw experimental $J_c(B, T, \theta)$ data is available online at RRI high-temperature superconducting (HTS) wire critical current database [56]). To perform the analysis, we assumed that REBCO film has the thickness of 1.7 μm, and all $\left|\overrightarrow{F_p}(J_c, T, \theta = 0°)\right|$ datasets were fitted to Eq. 28.

Deduced parameters are shown in Fig. 18. Overall, deduced the pining field enhancement factor, ε(*T*) (Figs. 13-17), in REBCO superconductors is varying within a range:

$$10 \leq \varepsilon(T)_{REBCO} \leq 80. \quad (42)$$

These values of $\varepsilon(T)_{REBCO}$ are in-times lower, than the $\varepsilon(T)_{NdFeAs(O,F)}$ in NdFeAs(O,F) thin film (Fig. 12).

The collective pinning peak is observable in all $\left|\overrightarrow{F_p}(J_c, T)\right|$ curves showed in Figs. 14,16,17. We should also note that there are a general similarity in the shape of $\left|\overrightarrow{F_p}(J_c, T = 4.2\ K)\right|$ measured in NdFeAs(O,F) film (Figure 10(a)) and (Gd,Y)$Ba_2Cu_3O_{7-\delta}$ + 15% $BaZrO_3$ film (Figure 15(c)), because both curve have a steep linear raise of $\left|\overrightarrow{F_p}(J_c, T = 4.2\ K)\right|$ at low applied magnetic fields.

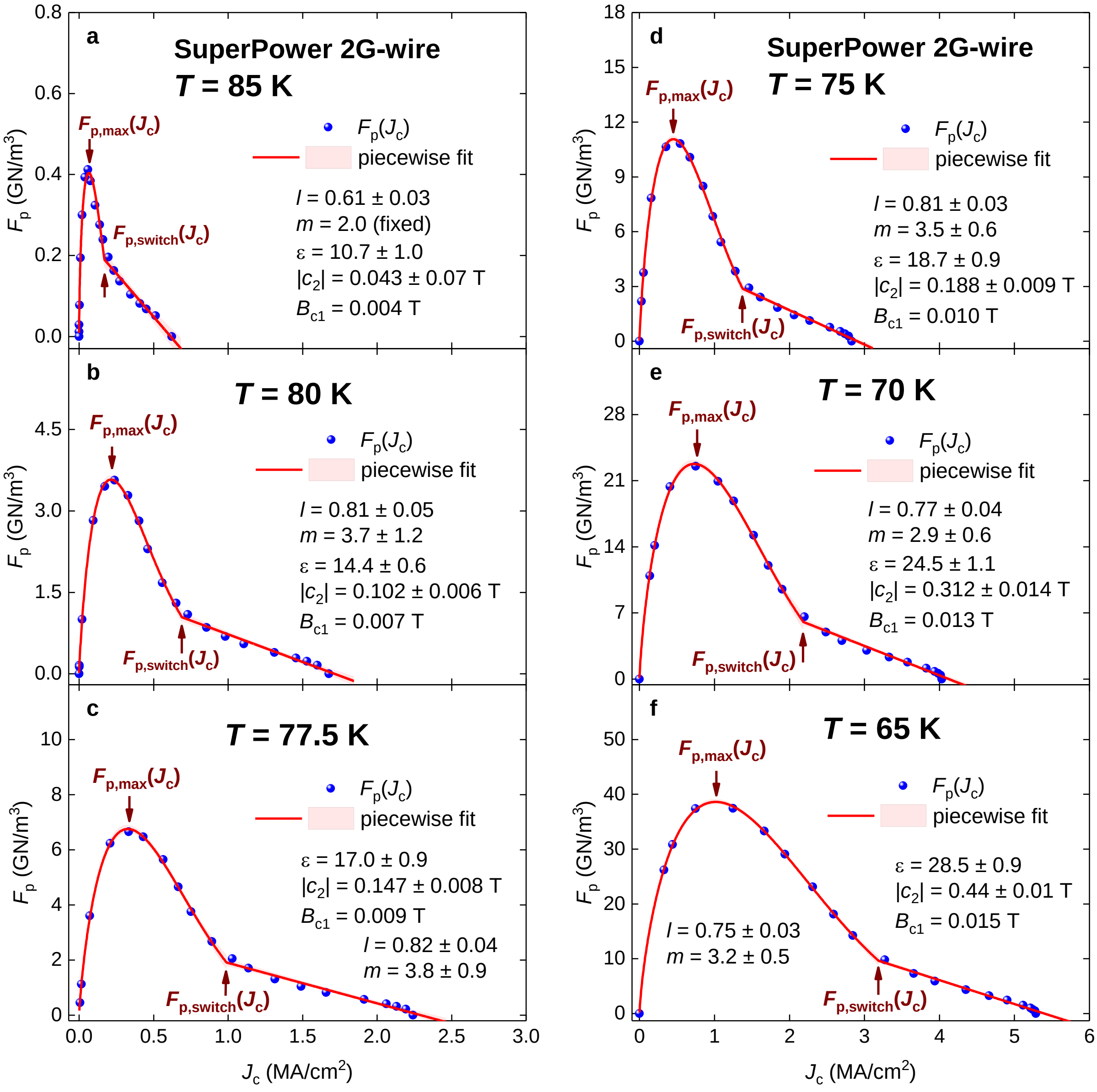

**Figure 16.** $|\overrightarrow{F_p}(J_c, T, \theta = 0°)|$ curves and data fits to Eq. 28 for SuperPower HTS 2G-wire measured at temperature range 85-65 K. Raw transport $J_c(B, T, \theta = 0°)$ data is freely available online [56]. Goodness of fit for all panels is better than $R$ = 0.998. 95% confidence bands are shown by shadow areas.

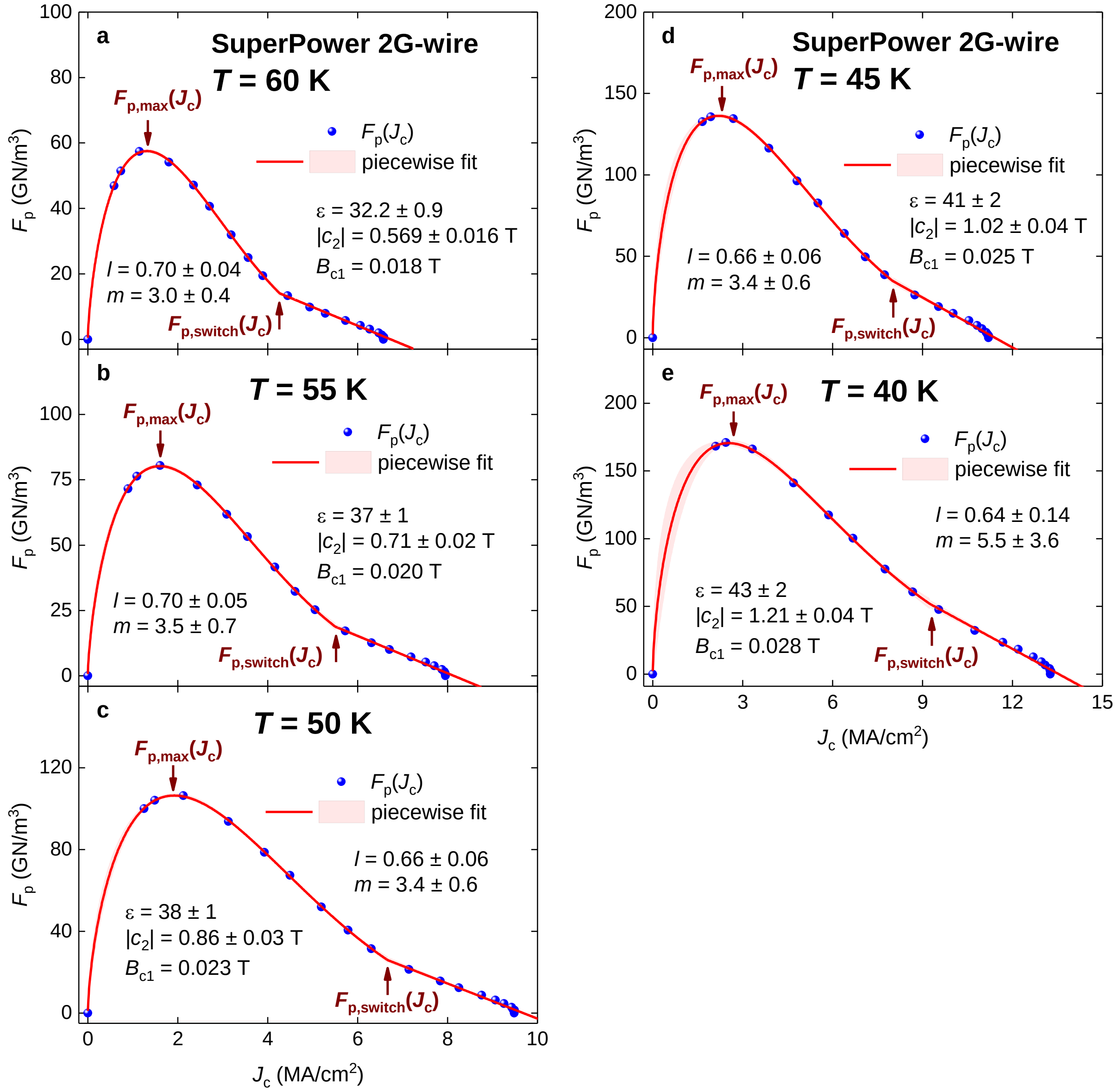

**Figure 17.** $\left|\overrightarrow{F_p}(J_c, T, \theta = 0°)\right|$ curves and data fits to Eq. 28 for SuperPower HTS 2G-wire measured at temperature range 60-40 K (a-f). Raw transport $J_c(B, T, \theta = 0°)$ data is freely available online [56]. Goodness of fit for all panels is better than $R$ = 0.998. 95% confidence bands are shown by shadow areas.

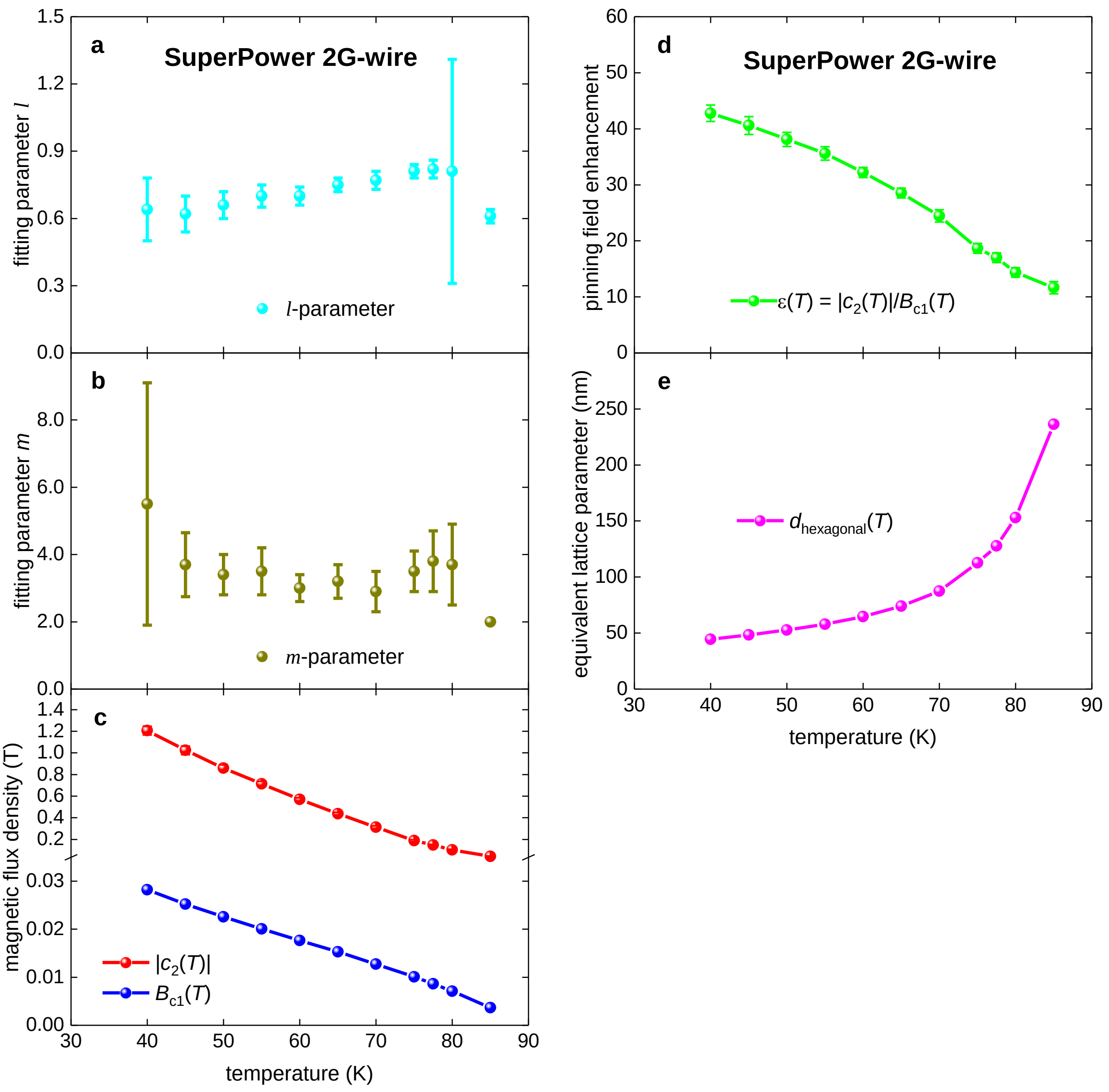

**Figure 18.** Deduced parameters from the fit of $\left|\overrightarrow{F_p}(J_c, T, \theta = 0°)\right|$ data to Eq. 28 for SuperPower HTS 2G-wire. **a** – $l(T)$ and ***b*** – $m(T)$ free-fitting parameters of Eq. 28; **c** – Deduced free-fitting parameter $|c_2(T)|$ (Eq. 13) and $B_{c1}(T)$ calculated by Eq. 21 for SuperPower wire for which raw transport $J_c(B, T, \theta = 0°)$ data is freely available online [52]. **d** – The pining field enhancement factor $\varepsilon(T)$ (Eq. 18) for the same film. **e** – Equivalent hexagonal lattice parameter $d_{hexagonal}(T)$ (Eq. 23) for the same film.

### 3.4. Near-room temperature superconductors (La,Y)$H_{10}$ and $YH_6$

Astonishing experimental discovery of near-room temperature superconductivity (NRTS) in highly compressed $H_3S$ by Drozdov *et al* [57], sparked world-wide research initiative in the hydrogen-rich superconductivity [58]. Recently, Semenok *et al* [59] predicted NRTS in ternary hydride (La,Y)$H_{10}$ by first-principle calculations and consequently synthesized this ternary hydride which exhibit $T_c$ up to 253 K, depends on applied pressure. Semenok *et al* [59] reported in-field transport critical current, $I_c(B,T)$, dataset for (La,Y)$H_{10}$ which exhibits zero resistance $T_c$ = 233 K at pressure $P$ = 186 K.

Here, we analysed $I_c(B,T=221\ K)$ and $R(T)$ data for the (La,Y)$H_{10}$ sample compressed at $P$ = 186 GPa (for which raw data is shown in Fig. 4 of Ref. 59). By assuming the sample thickness of 1 μm and the sample width of 20 μm, calculated $J_c(B,T=221\ K)$, $F_p(B,T=221\ K)$, and $F_p(J_c,T=221\ K)$ and showed these datasets in Fig. 19. We also added in these datasets experimental value for $B_{irr}(T=221\ K)=10\ T$ (which can be extracted from data showed in Fig. 4(b) in Ref. 59).

The most impressive deduced value for the (La,Y)$H_{10}$ sample is the pinning field enhancement factor, $\varepsilon(T=221\ K)$, which exceeds its counterpart in IBS NdFeAs(O,F) film (for calculations we utilized the Ginsburg-Landau parameter $\kappa(T)=90$, deduced for highly compressed $H_3S$ [60]):

$$|c_2(T=221\ K)| = 0.89 \pm 0.08\ T \quad (43)$$

$$B_{c1}(T=221\ K) = 0.9\ \pm 0.1\ mT \quad (44)$$

$$\varepsilon(T=221\ K) = 1000 \pm 100 \quad (45)$$

$$d_{hexagonal}(T=221\ K) = 52\ nm \quad (46)$$

In attempts to confirm this result, we analysed in-field transport critical current, $I_c(B,T)$, dataset for another NRTS superconductor $YH_6$, which was simultaneously discovered by Troyan *et al* [61] and Kong *et al* [62]. This superconductor exhibits zero resistance up to $T_c$ = 243 K depends on applied pressure.

Here, we analysed $I_c(B,T=189\ K)$ dataset for the $YH_6$ sample compressed at $P$ = 196 GPa (for which raw data is shown in Fig. 4(a) of Ref. 61). By assuming the sample thickness of 1 μm and the sample width of 20 μm, calculated $J_c(B,T=189\ K)$, $F_p(B,T=189\ K)$, and $F_p(J_c,T=189\ K)$ and showed these datasets in Fig. 20. For calculations we utilized the Ginsburg-Landau parameter $\kappa(T)=90$, deduced for highly compressed $H_3S$ [60].

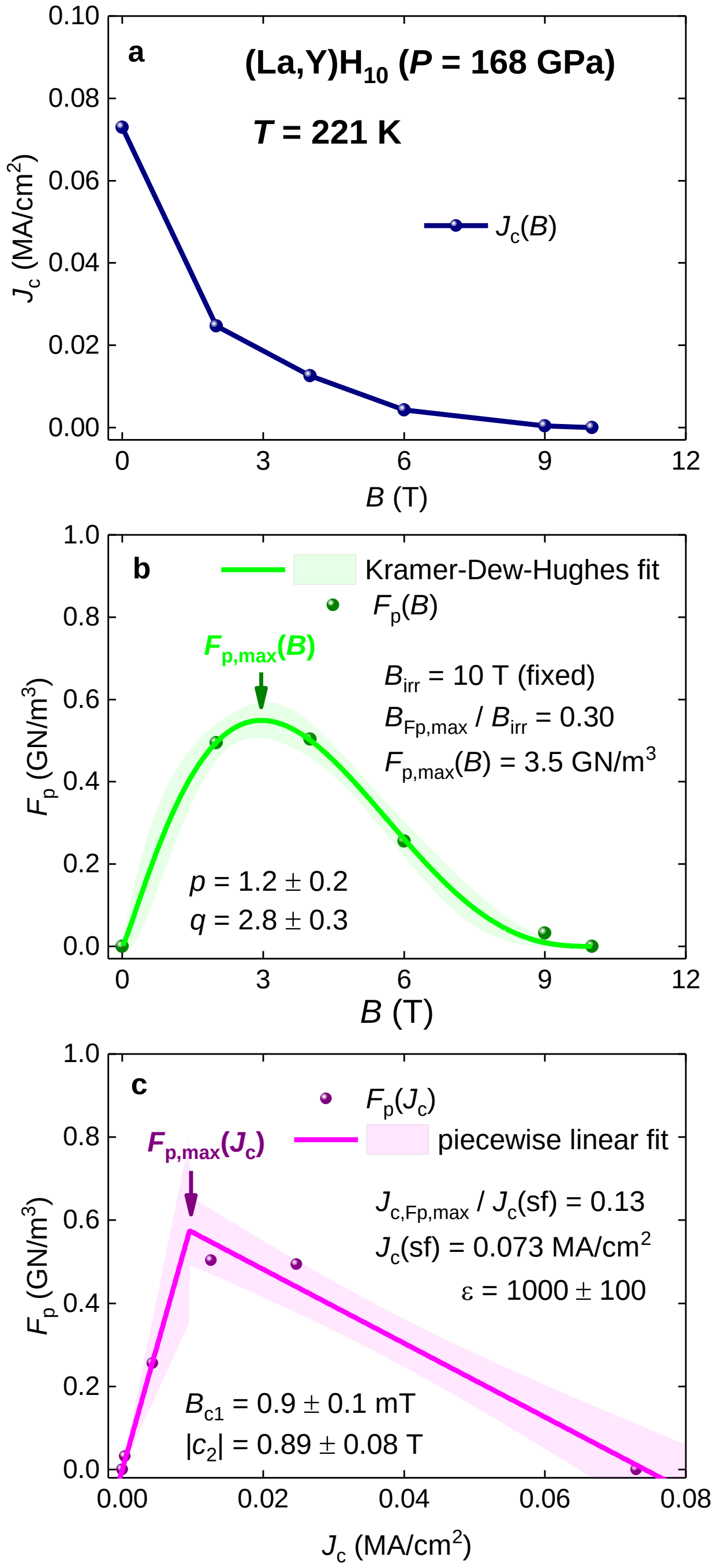

**Figure 19.** Projections of the $\left|\overrightarrow{F_p}(J_c, B, T = 221\ K)\right|$ curve for highly compressed (La,Y)$H_{10}$ ($P$ = 168 GPa) thin film into three orthogonal planes. Data fit for panel **b** and panel **c** are shown. **a** – the projection into the $F_p$=0 plane; **b** – the projection into the $J_c$=0 plane and data fit to Kramer-Dew-Hughes model (equation 3), $B_{irr}(T = 221\ K) = 10\ T$ (fixed value to experimentally observed value by Semenok *et al* [59]); and **c** – the projection into the $B$=0 plane and data fit to linear piecewise model (equation 13). Raw $I_c(B, T = 221\ K)$ data was reported by Semenok *et al* [59]. Assumed values are: the sample thickness is 1 μm, the sample width is 20 μm, the Ginsburg-Landau parameter is $\kappa(T) = 90$. 95% confidence bands are shown by shadow areas.

Deduced parameters for $YH_6$ ($P$ = 193 GPa) at $T$ = 189 K are:

$$|c_2(T = 189\ K)| = 1.02 \pm 0.16\ T \tag{47}$$

$$B_{c1}(T = 189\ K) = 1.1\ \pm 0.2\ mT \tag{48}$$

$$\varepsilon(T = 189\ K) = 950 \pm 200 \tag{49}$$

$$d_{hexagonal}(T = 189\ K) = 48\ nm \tag{50}$$

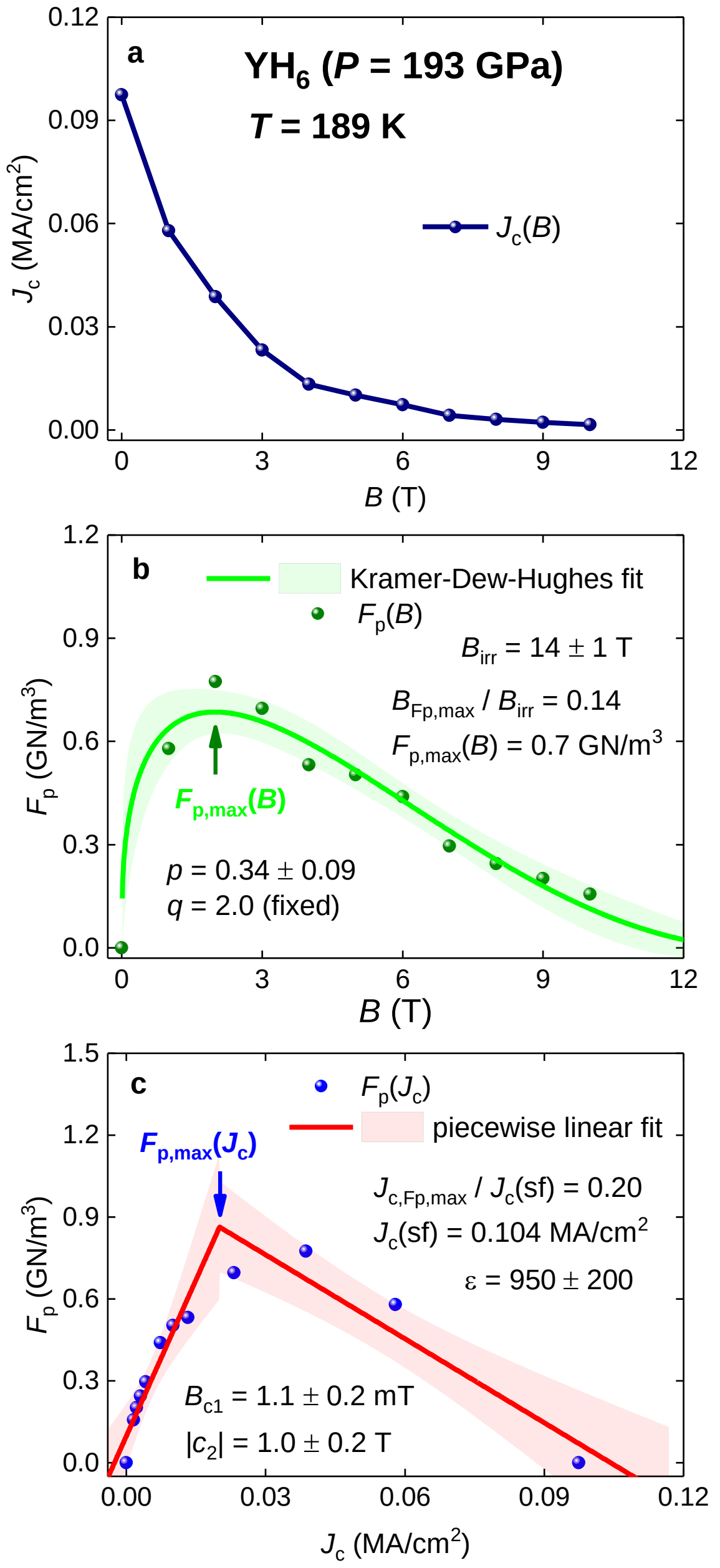

**Figure 19.** Projections of the $\left|\overrightarrow{F_p}(J_c, B, T = 189\ K)\right|$ curve for highly compressed $YH_6$($P$ = 193 GPa) thin film into three orthogonal planes. Data fit for panel **b** and panel **c** are shown. **a** – the projection into the $F_p$=0 plane; **b** – the projection into the $J_c$=0 plane and data fit to Kramer-Dew-Hughes model (equation 3), goodness of fit $R$ = 0.9635; and **c** – the projection into the $B$=0 plane and data fit to linear piecewise model (equation 13), goodness of fit $R$ = 0.9090. Raw $I_c(B, T = 189\ K)$ data was reported by Troyan *et al* [61]. Assumed values are: the sample thickness is 1 μm, the sample width is 20 μm, the Ginsburg-Landau parameter is $\kappa(T) = 90$. 95% confidence bands are shown by shadow areas.

Considering that in all studied materials $\varepsilon(T)$ is remaining to be the same order of magnitude within a wide temperature range, we can conclude that perhaps all NRTS have high values for the pinning field enhancement factor, $\varepsilon(T)$, and for the matching field (which we proposed should be attributed to the $|c_2(T)|$ field).

## 4. Conclusions

In conclusion, in this paper we proposed to consider the $F_p(J_c, T)$ data as a valuable source of experimental data which was (from our best knowledge) never considered in the literature. We analysed $F_p(J_c, T)$ datasets for thin films of $MgB_2$, NdFeAs(O,F), REBCO and highly compressed $(La,Y)H_{10}$ ($P$ = 186 GPa) and $YH_6$ ($P$ = 193 GPa). Our analysis leads us to a new definition for the matching field in superconducting materials.

### Data Availability Statement

The data that support the findings of this study are available from the author upon reasonable request.

### Conflicts of Interest

The authors have no conflicts to disclose.

### Acknowledgement

The author acknowledges financial support provided by the Ministry of Science and Higher Education of Russia (theme “Pressure” No. AAAA-A18-118020190104-3) and by Act 211 Government of the Russian Federation, Contract No. 02.A03.21.0006.